\documentclass[psfig]{mn2e}
\usepackage{psfig}

\def\sn{\hbox{S/N}}

\def\vsin{\hbox{$v \sin i$}} 
 
\def\kms{\hbox{km\,s$^{-1}$}}

\def\em{\it} 
 
\def\degr{\hbox{$^\circ$}} 
\def\rpd{\hbox{rad\,d$^{-1}$}}  
\def\mrpd{\hbox{mrad\,d$^{-1}$}}  
\def\omeq{\hbox{$\Omega_{\rm eq}$}}  
\def\dom{\hbox{$d\Omega$}}  
\def\kis{\hbox{$\chi^2$}}  
\def\kisr{\hbox{$\chi^2_{\rm r}$}}  
\def\drot{\hbox{differential rotation}}  
\def\hd{\hbox{HD~199178}}  
\def\hr{\hbox{HR~1099}}

\begin{document} 

\title[Magnetic field \& differential rotation of HD~199178] 
{Photospheric magnetic field and surface differential rotation of the FK~Com star HD~199178}
 
\makeatletter 
 
\def\newauthor{%
  \end{author@tabular}\par 
  \begin{author@tabular}[t]{@{}l@{}}} 
\makeatother 
  
\author[P.~Petit et al.] 
{\vspace{1.5mm} 
P.~Petit$^{1, 2}$, J.-F.~Donati$^2$, J.M.~Oliveira$^3$, M.~Auri\`ere$^2$, S.~Bagnulo$^4$, J.D.~Landstreet$^5$\\
{\hspace{-1mm}\vspace{1.5mm}\LARGE\rm F. Ligni\`eres$^2$, T. L\"uftinger$^6$, S.~Marsden$^2$, D. Mouillet$^2$, F. Paletou$^2$}\\
{\hspace{-1mm}\vspace{1.5mm}\LARGE\rm S.~Strasser$^7$, N.~Toqu\'e$^2$, G.A.~Wade$^8$}\\
$^1$Centro de Astrofisica da Universidade do Porto, rua das Estrelas, 4150-762 Porto, Portugal ({\tt petit@astro.up.pt})\\ 
$^2$Laboratoire d'Astrophysique, Observatoire Midi-Pyr\'en\'ees, 14 Av.\ E.~Belin, F--31400 Toulouse, France\\ ({\tt donati@ast.obs-mip.fr, auriere@ast.obs-mip.fr, francois.lignieres@obs-mip.fr, marsden@ast.obs-mip.fr}\\
{\tt mouillet@bagn.obs-mip.fr, toque@ast.obs-mip.fr})\\ 
$^3$School of chemistry and Physics, Keele University, Staffordshire ST5 5BG, UK ({\tt joana@astro.keele.ac.uk})\\
$^4$European Southern Observatory, Alonso de Cordova 3107, Vitacura, Santiago,
Chile ({\tt sbagnulo@eso.org})\\
$^5$Department of Physics and Astronomy, The University of Western Ontario, London, Ontario, Canada, N6G 3K7\\ ({\tt jlandstr@astro.uwo.ca})\\
$^6$Institut f\"ur Astronomie, Tuerkenschanzstrasse 17, A-1180 Wien, Austria ({\tt theresa@tycho.astro.univie.ac.at})\\
$^7$Department of Physics and Astronomy, University of Calgary, Calgary, AB T2N 1N4, Canada ({\tt strasser@ras.ucalgary.ca})\\
$^8$ Royal Military College of Canada, Department of Physics, P.O. Box 17000, Station "Forces", Kingston, Ontario, Canada, K7K 4B4\\ ({\tt Gregg.Wade@rmc.ca })\\}
\date{2003, MNRAS} 
\maketitle 
  
\begin{abstract}  
We present spectropolarimetric observations of the FK~Com star HD~199178 obtained between 1998 December and 2003 August at the T\'elescope Bernard Lyot (Observatoire du Pic du Midi, France). We report the detection of a photospheric magnetic field and reconstruct its distribution by means of Zeeman-Doppler Imaging. We observe large regions where the magnetic field is mainly azimuthal, suggesting that the dynamo processes generating the magnetic activity of \hd\ may be active very close to the stellar surface. We investigate the rapid evolution of surface brightness and magnetic structures from a continuous monitoring of the star over several weeks in 2002 and 2003. We report that significant changes occur in the distribution of cool spots and magnetic regions on typical timescales of the order of 2 weeks. Our spectropolarimetric observations also suggest that the surface of \hd\ is sheared by differential rotation, with a difference in rotation rate between equatorial and polar regions of the order of 1.5 times that of the Sun.

\end{abstract} 
 

\section{Introduction}  
\label{sect:introduction} 

Our understanding of the Solar dynamo has benefited from important progresses during the last decade, thanks to the combined breakthroughs of observational and modeling techniques. Helioseismological studies are now able to reveal the internal velocity field of the Sun (Schou et al. 1998) and recent MHD simulations give credence to the idea that the dynamo processes at the origin of the large-scale solar field are mainly operating within a thin interface layer separating the radiative core from the convective envelope of the Sun~: the tachocline. This region allows an efficient transformation of a seed poloidal field into a toroidal component as well as a storage of the field on a timescale of the order of the solar magnetic cycle (Rempel et al. 2000). The poloidal component of the field is believed to be regenerated (with opposite polarity) through the action of the Coriolis force, by means of a still mostly unknown mechanism called ``$\alpha$-effect'', which may also be mostly efficient within the tachocline (Dikpati \& Gilman 2001). 

The magnetic field observed at the photospheric level of the Sun was first detected in sunspots (Hale 1908) which are the largest magnetic structures of the solar surface (though only covering about $10^{-4}$ of the photosphere at solar maximum), featuring mostly radially oriented field lines. Magnetic elements are now observed at much smaller scales in the so-called quiet photosphere (see Solanki 2001 for a review of the properties of such magnetic regions). The interplay, during the solar cycle, between the small magnetic structures and the large-scale field has not yet been understood, whereas small magnetic elements store a significant part of the magnetic energy of the photosphere, and despite the fact the distribution of some of these structures has a temporal evolution correlated to the solar cycle. Some studies also suggest that a turbulent dynamo, disconnected from the large-scale solar dynamo, may efficiently generate small-scale magnetic structures (Cattaneo 1999). 

Surprisingly, these properties of the solar photospheric field are far from the general features observed on other active stars, at least among the fast rotator group. The surface magnetic field of several fast rotating stars has been mapped by means of Zeeman-Doppler Imaging (thereafter ZDI, Semel 1989). For these extremely active objects, a field is generally observed at basically all locations of the photosphere (see e.g. Donati et al. 2003a), in the form of discrete structures that present an homogeneous orientation of field lines and cover a significant fraction of the stellar surface. On some objects, the surface field is dominated by an azimuthal component distributed in arcs or even in complete rings encircling the rotation axis at different latitudes. These observations suggest that the dynamo operating within these objects is very different from that of the Sun, since the azimuthal component of the solar field is supposed to be deeply buried at the base of the convective envelope. Its presence at the photospheric level on fast rotators therefore suggests that their dynamo may be active very close to the surface.

The observing effort engaged during the last decade to investigate, by means of high-resolution spectropolarimetry, the magnetic activity of active late-type fast rotators has been concentrated on several pre-main sequence stars and on the evolved close binary system HR~1099. The present study focuses on \hd, a member of the small FK~Com group. Objects belonging to this class are single fast rotating late-type giants displaying signs of a strong magnetic activity. Their short rotation periods (of order of a few days) suggest that FK~Com stars have undergone very unusual mechanisms during their evolution. In particular, it was proposed by Bopp \& Stencel (1981) that they may result from the coalescence of contact binaries. Long-term photometric studies of FK~Com itself (Jetsu et al. 1991, Korhonen et al. 2002) report that its cool spot distribution is asymmetric (a so-called ``active longitude'' dominating the spot activity), with occasional 180\degr\ shifts of the active longitude (a behavior dubbed ``flip-flop'') resulting in a 6.5 years cycle. The distribution of cool spots of \hd\ has also been investigated by means of Doppler imaging (Strassmeier et al. 1999, Hackman et al. 2001). 

The aim of the present study is to enrich our knowledge of this object by reconstructing its magnetic topology altogether with its spot distribution, in a twofold purpose. We first plan to compare the general properties of the photospheric magnetic field of this evolved star with the magnetic topologies mapped for younger objects. We also intend to check whether active longitudes reported by Hackman et al. (2001) are also observed in our own maps of the cool spot distribution and test whether they have a counterpart in the magnetic images. 

We first summarize the series of spectropolarimetric observations used for this study and describe the modeling procedures employed to reconstruct the magnetic topology of the star at every observing epoch. We then present the main characteristics of its surface brightness and magnetic structures and study their evolution on different timescales (from a few weeks to several years), including an analysis of the surface \drot. We finally summarize the main results and outline the informations they provide on the nature of the underlying dynamo processes.  
  
\section{Observations, data reduction and imaging procedure}
\label{sect:observations}

\subsection{Observations}

All the spectropolarimetric data presented in this article were obtained with the MuSiCoS spectrograph (Baudrand \& B\"ohm 1992) fiber fed by its Cassegrain-mounted polarimetric module (Donati et al. 1999). The data reduction, performed with ESpRIT (Donati et al. 1997), is similar to that described by Petit et al. (2003) for HR~1099, including the additional wavelength calibration using telluric lines. The data sets (listed in Tab. \ref{tab:journal} and \ref{tab:journal2}) were obtained between 1998 Dec. and 2003 Aug., yielding a total number of 380 brightness spectra and 94 circularly polarized spectra. 

Least-Square Deconvolution (thereafter LSD, Donati et al. 1997) was employed to perform a simultaneous extraction of the signal from all photospheric spectral features of the echelograms. A line mask corresponding to a G5 spectral type (Strassmeier et al. 1999) yielded a multiplex gain of the order of 30 for the Stokes V profiles. The \sn\ of LSD Stokes I profiles is limited to about 1100 at best, indicating that the convolution model underlying LSD is not adapted above this accuracy level. Depending on data quality, the multiplex gain undergoes small fluctuations, so that raw spectra of equal \sn\ can produce LSD profiles with slightly different noise levels. 

The data quality of different observing periods is very uneven (due to weather conditions), so that the \sn\ of LSD profiles ranges from about 3,600 (in average in 2001 Jul.) to more than 7,000 on 2003 Jul. 31 (in optimal seeing and transparency conditions). The large data sets of summers 2002 and 2003 are split into several subsets (separated by blank lines in Tab. \ref{tab:journal} and \ref{tab:journal2}). Each subset is used to reconstruct an individual brightness and magnetic image of the star (Sect. \ref{sect:procedure}), ensuring that each map is computed from observations obtained over a time-span consistent with the typical short-term evolution of photospheric structures (Sect. \ref{sect:changes}).   

\begin{table*}
\caption[]{Journal of observations from 1998 Dec. to 2002 Jul.. Each line corresponds to a full polarization cycle. Columns 2 and 3 list the date and hour of observation (first and last exposure of the cycle). Column 4 contains the number of unpolarized/polarized exposures. Column 5 lists the total exposure time of each Stokes I individual sub-exposure. We also list the \sn\ ratios (per 4 \kms\ velocity bins) of the unpolarized and polarized spectra (in columns 6 and 8 respectively, with minimum and maximum values in the sequence) and in the associated mean LSD profiles (columns 7 and 9). The multiplex gain between the raw polarized spectra and the mean Stokes V profiles is reported in the last column.}
\begin{tabular}{cccccccccc}
\hline
Date  & JD           & UT & nexp & t$_{\rm exp}$ & \sn & \sn          & \sn  & \sn          & multiplex gain \\
      & (+2,450,000) & (hh$:$mm$:$ss)   &      & (sec.)        &  I  & I$_{\rm LSD}$ & V    & V$_{\rm LSD}$ & V \\
\hline
1998 Dec 5 & 1153.29/1153.32 & 18:58:31/19:34:00 & 4/1 & 600 & 90/100 & 1053/1072 & 180 & 5509 & 31\\

1999 Jan 14 & 1193.25/1193.27 & 18:00:57/18:34:16 & 4/1 & 600 & 80/90 & 759/855 & 170 & 4690 & 28\\

1999 Jan 18 & 1197.25/1197.27 & 18:00:59/18:34:38 & 4/1 & 600 & 100/100 & 1074/1081 & 190 & 4997 & 26\\

1999 Jan 19 & 1198.25/1198.27 & 18:02:20/18:35:28 & 4/1 & 600 & 90/100 & 970/1018 & 170 & 4599 & 27\\
\hline
2001 Jul 3 & 2094.44/2094.49 & 22:38:11/23:51:16 & 5/1 & 900 & 60/100 & 966/1066 & 190 & 5873 & 31\\

2001 Jul 7 & 2098.45/2098.48 & 22:41:03/23:34:41 & 4/1 & 145/900 & 30/90 & 686/1069 & 150 & 2908 & 19\\

2001 Jul 18 & 2109.46/2109.49 & 23:06:23/23:41:08 & 3/1 & 900 & 20/30 & 435/556 & 34 & 768 & 23\\

2001 Jul 25 & 2115.53/2115.56 & 00:37:14/01:24:35 & 4/1 & 900 & 50/60 & 838/937 & 110 & 2824 & 26\\

2001 Jul 26 & 2117.41/2117.44 & 21:46:48/22:34:10 & 4/1 & 900 & 100/100 & 1062/1091 & 200 & 5820 & 29\\
\hline
2001 Dec 1 & 2245.26/2245.29 & 18:16:05/19:03:26 & 4/1 & 900 & 70/80 & 897/1038 & 140 & 3895 & 28\\

2001 Dec 2 & 2246.23/2246.26 & 17:33:05/18:20:26 & 4/1 & 900 & 100/100 & 1053/1070 & 190 & 5556 & 29\\

2001 Dec 7 & 2251.23/2251.26 & 17:31:59/18:19:20 & 4/1 & 900 & 100/100 & 1053/1066 & 190 & 5463 & 29\\

2001 Dec 8 & 2252.23/2252.27 & 17:35:20/18:22:41 & 4/1 & 900 & 90/100 & 1063/1073 & 180 & 5325 & 30\\

2001 Dec 9 & 2253.23/2253.27 & 17:36:42/18:24:04 & 4/1 & 900 & 100/110 & 1072/1078 & 200 & 6064 & 30\\

2001 Dec 10 & 2254.24/2254.27 & 17:43:08/18:30:30 & 4/1 & 900 & 80/100 & 1024/1073 & 170 & 5061 & 30\\

2001 Dec 11 & 2255.24/2255.27 & 17:42:24/18:29:45 & 4/1 & 900 & 100/110 & 1056/1071 & 210 & 6206 & 30\\

2001 Dec 12 & 2256.24/2256.27 & 17:38:59/18:26:21 & 4/1 & 900 & 100/110 & 1065/1088 & 210 & 6310 & 30\\

2001 Dec 13 & 2257.24/2257.27 & 17:38:55/18:26:16 & 4/1 & 900 & 170/180 & 754/882 & 220 & 6844 & 31\\

2001 Dec 16 & 2260.24/2260.27 & 17:39:23/18:26:44 & 4/1 & 900 & 90/100 & 1047/1157 & 180 & 5087 & 28\\
\hline

2002 Jun 11 & 2437.47/2437.51 & 23:10:02/23:59:11 & 4/1 & 900 & 60/80 & 982/1045 & 130 & 3486 & 27\\

2002 Jun 12 & 2438.48/2438.51 & 23:31:33/00:18:55 & 4/1 & 900 & 40/60 & 827/941 & 98 & 2485 & 25\\

2002 Jun 13 & 2439.49/2439.52 & 23:39:48/00:27:12 & 4/1 & 900 & 70/90 & 909/1054 & 160 & 4497 & 28\\

2002 Jun 15 & 2441.47/2441.51 & 23:16:34/00:07:37 & 4/1 & 900 & 80/90 & 1043/1075 & 170 & 5081 & 30\\

2002 Jun 18 & 2444.41/2444.44 & 21:49:48/22:37:09 & 4/1 & 900 & 60/90 & 933/1064 & 130 & 3801 & 29\\

2002 Jun 22 & 2448.45/2448.48 & 22:41:35/23:28:56 & 4/1 & 900 & 100/110 & 1090/1111 & 210 & 6126 & 29\\

2002 Jun 25 & 2451.40/2451.44 & 21:40:20/22:27:43 & 4/1 & 900 & 90/100 & 1062/1090 & 170 & 4993 & 29\\

 &&&&&&&&&\\
2002 Jun 28 & 2454.43/2454.46 & 22:20:47/23:08:00 & 4/1 & 900 & 80/100 & 1037/1109 & 180 & 4842 & 27\\

2002 Jun 29 & 2454.57/2454.60 & 01:39:51/02:27:12 & 4/1 & 900 & 40/100 & 814/1090 & 160 & 3485 & 22\\

2002 Jun 29 & 2455.42/2455.45 & 22:05:59/22:53:20 & 4/1 & 900 & 70/90 & 991/1055 & 160 & 4474 & 28\\

2002 Jun 30 & 2455.57/2455.60 & 01:42:08/02:29:29 & 4/1 & 900 & 80/100 & 1006/1086 & 190 & 5378 & 28\\

2002 Jul 1 & 2457.50/2456.53 & 00:00:31/00:47:53 & 4/1 & 900 & 70/80 & 937/1036 & 130 & 3625 & 28\\

2002 Jul 2 & 2458.43/2458.46 & 22:18:02/23:05:23 & 4/1 & 900 & 70/70 & 1013/1031 & 130 & 3728 & 29\\

2002 Jul 3 & 2458.53/2458.56 & 00:42:02/01:29:23 & 4/1 & 900 & 60/70 & 968/1004 & 120 & 3199 & 27\\

2002 Jul 4 & 2459.55/2459.59 & 01:15:22/02:02:43 & 4/1 & 900 & 60/70 & 909/1002 & 120 & 3550 & 30\\

2002 Jul 4 & 2460.42/2460.45 & 22:02:50/22:50:11 & 4/1 & 900 & 50/60 & 924/995 & 110 & 2971 & 27\\

2002 Jul 5 & 2460.52/2460.55 & 00:24:40/01:12:01 & 4/1 & 900 & 60/70 & 992/1024 & 130 & 3514 & 27\\

2002 Jul 5 & 2460.62/2460.65 & 02:55:48/03:43:09 & 4/1 & 900 & 80/100 & 1057/1083 & 160 & 4769 & 30\\

2002 Jul 11 & 2467.41/2467.44 & 21:45:13/22:32:38 & 4/1 & 900 & 70/80 & 989/1045 & 140 & 3653 & 26\\

 &&&&&&&&&\\
2002 Jul 12 & 2467.54 & 00:56:31 & 1/0 & 900 & 80 & 1033 & -- & -- & --\\
2002 Jul 17 & 2473.41/2473.44 & 21:44:56/22:32:12 & 4/1 & 900 & 140/150 & 715/817 & 300 & 4215 & 14\\

2002 Jul 17 & 2473.49/ 2473.52 & 23:44:49/00:32:04 & 4/1 & 900 & 100/110 & 630/675 & 210 & 3293 & 16\\

2002 Jul 18 & 2474.44/ 2474.48 & 22:40:41/ 23:27:58 & 4/1 & 900 & 30/50 & 589/838 & 76 & 1833 & 24\\

2002 Jul 19 & 2474.58/2474.61 & 01:51:52/02:39:8 & 4/1 & 900 & 40/60 & 765/974 & 94 & 2384 & 25\\

2002 Jul 19 & 2475.40/2475.44 & 21:42:56/22:30:12 & 4/1 & 900 & 30/50 & 660/828 & 82 & 1934 & 24\\

2002 Jul 19 & 2475.45/2475.48 & 22:50:32/23:37:47 & 3/0 & 900 & 30/50 & 509/878 & -- & -- & -- \\

2002 Jul 21 & 2477.49/2477.52 & 23:44:47/00:32:26 & 4/1 & 900 & 70/70 & 945/970 & 120 & 3225 & 27\\

2002 Jul 22 & 2477.59/2477.62 & 02:04:22/02:52:02 & 4/1 & 900 & 70/80 & 998/1052 & 140 & 3742 & 27\\

2002 Jul 26 & 2482.43/2482.47 & 22:22:46/23:10:01 & 4/1 & 900 & 80/80 & 1040/1045 & 150 & 4034 & 27\\

2002 Jul 27 & 2482.53/2482.56 & 00:41:53/01:29:08 & 4/1 & 900 & 70/80 & 972/1048 & 150 & 4102 & 27\\

2002 Jul 27 & 2482.63/2482.67 & 03:14:15/04:01:30 & 4/1 & 900 & 100/100 & 1062/1091 & 200 & 5941 & 30\\

2002 Jul 27 & 2483.43/2483.46 & 22:15:20/23:02:35 & 4/1 & 900 & 100/100 & 1038/1092 & 190 & 5617 & 30\\

2002 Jul 28 & 2483.52/2483.56 & 00:34:07/01:21:23 & 4/1 & 900 & 90/90 & 1051/1075 & 180 & 5077 & 28\\

2002 Jul 28 & 2483.63/2483.66 & 03:5:23/03:52:39 & 4/1 & 900 & 90/100 & 1041/1083 & 180 & 4980 & 28\\

\hline
\end{tabular}
\label{tab:journal}
\end{table*}

\begin{table*}
\caption[]{Same as Table \ref{tab:journal} for summer 2003 observations.}\begin{tabular}{cccccccccc}
\hline
Date  & JD           & UT & nexp & t$_{\rm exp}$ & \sn & \sn          & \sn  & \sn          & multiplex gain \\
      & (+2,450,000) & (hh$:$mm$:$ss)   &      & (sec.)        &  I  & I$_{\rm LSD}$ & V    & V$_{\rm LSD}$ & V \\
\hline
2003 Jun 27 & 2817.58/2817.61 & 01:55:40/02:38:20 & 4/1 & 800 & 100/100 & 1095/1101 & 200 & 6053 & 30 \\
2003 Jun 29 & 2819.59/2819.62 & 02:10:10/02:52:50 & 4/1 & 800 & 70/80 & 1031/1042 & 140 & 3953 & 28 \\
2003 Jul 4 & 2825.48/2825.51 & 23:34:01/00:16:42 & 4/1 & 800 & 90/100 & 1077/1104 & 170 & 4787 & 28 \\
2003 Jul 6 & 2826.59/2826.62 & 02:03:18/02:45:58 & 4/1 & 800 & 60/90 & 930/1074 & 140 & 3825 & 27 \\
2003 Jul 7 & 2827.53/2827.56 & 00:37:50/01:20:30 & 4/1 & 800 & 100/100 & 1076/1083 & 190 & 5528 & 29 \\
2003 Jul 8 & 2828.50/2828.53 & 24:06:10/00:48:50 & 4/1 & 800 & 70/80 & 993/1037 & 130 & 3773 & 29 \\
2003 Jul 10 & 2830.55/2830.58 & 01:11:40/01:54:20 & 4/1 & 800 & 90/100 & 1010/1086 & 180 & 5382 & 30\\
2003 Jul 11 & 2831.54/2831.57 & 00:59:25/01:42:06 & 4/1 & 800 & 60/90 & 892/1030 & 130 & 3749 & 29\\
 &&&&&&&&&\\
2003 Jul 12 & 2832.58/2832.61 & 01:52:08/02:34:48 & 4/1 & 800 & 90/110 & 1080/1090 & 190 & 5550 & 29\\
2003 Jul 12 & 2833.46/2833.49 & 23:06:00/23:48:42 & 4/1 & 800 & 80/90 & 1051/1058 & 160 & 4465 & 28\\
2003 Jul 15 & 2835.57/2835.60 & 01:37:13/02:19:29 & 4/1 & 800 & 70/90 & 1054/1081 & 160 & 4610 & 29\\
2003 Jul 17 & 2837.57/2837.60 & 01:39:30/02:21:51 & 4/1 & 800 & 50/70 & 905/1035 & 120 & 3383 & 28\\
2003 Jul 17 & 2838.47/2838.50 & 23:18:50/24:01:00 & 4/1 & 800 & 70/80 & 1027/1063 & 150 & 4281 & 29\\
2003 Jul 18 & 2839.46/2839.49 & 23:04:30/23:46:46 & 4/1 & 800 & 80/90 & 1016/1065 & 160 & 4716 & 29\\
2003 Jul 19 & 2840.45/2840.48 & 22:53:17/23:35:32 & 4/1 & 800 & 90/100 & 1034/1081 & 180 & 5268 & 29\\
2003 Jul 20 & 2841.46/2841.49 & 23:07:40/23:49:55 & 4/1 & 800 & 80/80 & 1039/1058 & 150 & 4310 & 29\\
2003 Jul 21 & 2841.59/2841.62 & 02:12:15/02:54:31 & 4/1 & 800 & 70/80 & 973/1032 & 130 & 3893 & 30\\
2003 Jul 22 & 2842.55/2842.58 & 01:16:28/01:58:44 & 4/1 & 800 & 100/100 & 1102/1093 & 90 & 5607 & 29 \\
2003 Jul 22 & 2843.46/2843.48 & 22:55:15/23:37:31 & 4/1 & 800 & 90/100 & 1061/1074 & 180 & 5412 & 30 \\
2003 Jul 23 & 2843.55/2843.58 & 01:13:27/01:55:44 & 4/1 & 800 & 100/100 & 1074/1088 & 190 & 5540 & 29\\
2003 Jul 23 & 2844.45/2844.48 & 22:46:45/23:29:01 & 4/1 & 800 & 100/100 & 1078/1096 & 190 & 5686 & 30\\
2003 Jul 24 & 2844.55/2844.58 & 01:08:41/01:50:57 & 4/1 & 800 & 10/90 & 268/1072 & 110 & 1162 & 11 \\
2003 Jul 24 & 2845.47/2845.50 & 23:17:32/23:59:48 & 4/1 & 800 & 90/90 & 1068/1084 & 170 & 4988 & 29\\
2003 Jul 25 & 2845.59/2845.62 & 02:06:23/02:48:39 & 4/1 & 800 & 80/80 & 1033/1051 & 150 & 4315 & 29\\
2003 Jul 25 & 2846.47/2846.50 & 23:13:19/23:55:35 & 4/1 & 800 & 40/80 & 740/1034 & 110 & 2806 & 25\\
 &&&&&&&&&\\
2003 Jul 27 & 2847.60/2847.63 & 02:21:46/03:04:00 & 4/1 & 800 & 60/90 & 979/1069 & 140 & 4196 & 30\\
2003 Jul 27 & 2848.45/2848.48 & 22:51:13/23:33:28 & 4/1 & 800 & 70/80 & 1025/1053 & 140 & 4130 & 29\\
2003 Jul 28 & 2848.58/2848.61 & 01:57:30/02:39:46 & 4/1 & 800 & 80/90 & 1049/1077 & 160 & 4784 & 30\\
2003 Jul 29 & 2849.54/2849.57 & 01:00:45/01:43:01 & 4/1 & 800 & 110/110 & 1103/1107 & 220 & 6718 & 30\\
2003 Jul 29 & 2850.41/2850.44 & 21:47:58/22:30:14 & 4/1 & 800 & 100/110 & 1014/1118 & 210 & 6399 & 30\\
2003 Jul 30 & 2850.58/2850.61 & 02:01:36/02:43:51 & 4/1 & 800 & 100/110 & 1050/1104 & 210 & 6405 & 30\\
2003 Jul 30 & 2851.41/2851.44 & 21:45:27/22:27:43 & 4/1 & 800 & 90/90 & 1022/1087 & 180 & 5305 & 29\\
2003 Jul 31 & 2851.53/2851.56 & 00:42:03/01:29:18 & 4/1 & 900 & 100/110 & 1081/1113 & 200 & 6003 & 30\\
2003 Jul 31 & 2852.41/2852.44 & 21:54:40/22:36:56 & 4/1 & 800 & 110/120 & 1041/1108 & 230 & 7090 & 31\\
2003 Aug 1 & 2852.59/2852.62 & 02:10:05/02:52:21 & 4/1 & 800 & 110/110 & 987/1120 & 220 & 6669 & 30\\
2003 Aug 3 & 2854.54/2854.57 & 00:52:43/01:34:59 & 4/1 & 800 & 110/110 & 959/1114 & 210 & 6483 & 31\\
2003 Aug 3 & 2854.58/2854.61 & 01:49:01/02:31:17 & 4/1 & 800 & 110/110 & 980/1096 & 220 & 6548 & 30\\
2003 Aug 4 & 2855.55/2855.58 & 01:09:28/01:51:44 & 4/1 & 800 & 90/90 & 971/1020 & 160 & 4419 & 28\\
2003 Aug 4 & 2855.59/2855.62 & 02:05:47/02:48:03 & 4/1 & 800 & 80/90 & 910/984 & 170 & 4831 & 28\\
2003 Aug 4 & 2856.46/2856.49 & 23:08:58/23:51:14 & 4/1 & 800 & 100/100 & 1084/1107 & 200 & 5936 & 30\\
2003 Aug 5 & 2856.60/2856.63 & 02:30:45/03:13:01 & 4/1 & 800 & 80/100 & 936/1035 & 180 & 5160 & 29\\
2003 Aug 5 & 2857.48/2857.51 & 23:34:09/00:16:25 & 4/1 & 800 & 100/110 & 1002/1088 & 200 & 6033 & 30\\
2003 Aug 6 & 2857.56/2857.59 & 01:24:59/02:07:15 & 4/1 & 800 & 100/110 & 1061/1094 & 210 & 6398 & 30\\

\hline
\end{tabular}
\label{tab:journal2}
\end{table*}

\subsection{Imaging procedure}
\label{sect:procedure}

All magnetic images described hereafter are obtained with the ZDI code developed by Brown et al. (1991) and Donati \& Brown (1997), following the maximum entropy image reconstruction algorithm of Skilling \& Bryan (1984). This imaging procedure was tested for various stellar parameters and observing conditions by Donati \& Brown (1997), from a series of numerical simulations. They demonstrated that the orientation of field lines within magnetic regions can be accurately reconstructed for noise levels similar to that available in the present study and for inclination angles of the rotation axis similar to that derived later in this section for \hd. However, in case of images reconstructed with data presenting an incomplete phase sampling of the star, only a partial reconstruction of the magnetic field is achieved, containing radial/meridional field regions closest to the observed longitudes and azimuthal field structures located about 0.1 rotation cycle away from the observed longitudes. With sparse phase coverage, the reconstructed position of active regions located next to phase gaps can also be slightly shifted toward higher latitudes and unobserved phases (Petit et al. 2002). Comparisons between ZDI images of HR~1099 reported by Donati et al. (2003a) and Petit et al. (2003) demonstrate however that maps of the same object obtained with simultaneous observations, but with different instrumental setups, are very consistent despite very different noise levels and phase sampling. 

We model the photospheric brightness inhomogeneities by means of the two component description of Cameron et al. (1992). The average intrinsic profile used for computing brightness images is a synthetic Gaussian line reproducing the characteristics of a MuSiCoS LSD Stokes I profile of the K0 star $\beta$~Gem. This option was adopted according to the results of Unruh \& Cameron (1995), who demonstrated that Doppler images reconstructed from a Gaussian line were almost indistinguishable from that obtained using a standard star. The template profile was scaled by a factor 0.5 and 1, for the spotted areas and the quiet photosphere respectively, to mimic the different LSD signatures expected from these two regions (Donati \& Cameron 1997). The adopted temperatures are 5500~K and 4000~K, respectively for the quiet photosphere and the spotted regions, both values being close to that determined by O'Neal et al. (1998) from TiO band modeling.

The large volume of data we collected for the present study allows us to obtain estimates of the imaging parameters associated with \hd. To estimate the radial velocity $v_{\rm rad}$, we reconstruct several images from the same data set, tuning the value of $v_{\rm rad}$ and choosing the one that minimizes the information content of the image (following Cameron \& Unruh 1994). In order to estimate the projected rotational velocity \vsin, we prefer to adopt the more reliable estimate provided by a classical fit to the data (Donati et al. 2003a), from which we can obtain a better adjustment of the profiles wings, at the cost of a slightly higher spot (or magnetic field) coverage. The values we obtain for $v_{\rm rad}$ and \vsin\ are respectively equal to $-28.5 \pm 0.5~\kms$ and $70\pm1~\kms$, in good agreement with previous estimates of Strassmeier et al. (1999).

The rotation period (or the rotation law in the case of a differentially rotating surface) can be estimated in a similar way as $v_{\rm rad}$, but we leave the detailed analysis of surface \drot\ for Sect. \ref{sect:diffrot}. All the images presented hereafter are computed assuming that the surface undergoes differential rotation during data collection. We assume in the reconstruction process a surface rotation law of the type:

\begin{equation}
\Omega(l) = \omeq - \dom \sin^2 l
\label{eq:diffrot}
\end{equation}

\noindent where $\Omega(l)$ is the rotation rate at latitude $l$, \omeq\ the rotation rate of the equator (set to 1.934 \rpd, see Sect. \ref{sect:diffrot}) and \dom\ the difference in rotation rate between the pole and the equator (set to 66 \mrpd\ by default). We assume a mean rotation period equal to 3.3~d (1.904 \rpd) for calculating the rotational phases. Given the differential rotation parameters assumed for the star, this period is the one we expect at latitude 42\degr\ (rotation periods of the stellar surface varying from 3.25~d at the equator to 3.36~d in the polar region). The adopted ephemeris is~:

\begin{equation}
JD = 2451150.675 + 3.3~\phi
\label{eq:ephemeris}
\end{equation}

\noindent where $\phi$ is the rotational phase and $JD$ the Julian date. Considering this $3.3\pm0.05$~d rotation period and using the \vsin\ value derived above, we estimate the stellar radius to be $R.{\rm sin}i = (4.5 \pm 0.1)~R_{\odot}$. This value is compatible with a previous estimate of Hackman et al. (2001, based on Hipparcos parallax measurements) that yields a stellar radius of $4~R_{\odot} < R < 5~R_{\odot}$). However, high values of $i$ are obviously required to stay consistent with the Hipparcos radius. Considering several Doppler images reconstructed with different values of the inclination, the highest possible value of $i$ is about 50\degr, since higher inclination angles produce obviously spurious structures on the images. For this inclination the value of the stellar radius is roughly 5.8~$R_{\odot}$, higher than that proposed by Hackman et al.. However, as emphasized by these authors and by Strassmeier et al. (1999), the Hipparcos-based estimate critically depends on the physical models used to describe such a post-main sequence star. Given the rather unusual physical properties of \hd\ (among which its high rotation rate and its uncertain evolution on the main sequence, since this single star may come from the coalescence of a close binary system), this kind of discrepancy is not surprising, therefore we choose to use our estimate $i=50$\degr. 

To summarize, stellar parameters adopted in the imaging process are $i=50$\degr, $v_{\rm rad} = -28.5~\kms$, \vsin$=70~\kms$, \omeq$=1.934~\rpd$ and \dom$=66~\mrpd$.

\section{Reconstructed images}

Nine brightness and eight magnetic images were reconstructed from the data sets described in Sect. \ref{sect:observations}, corresponding to epochs 1999.02, 2001.54, 2001.97, 2002.46, 2002.50, 2002.56, 2003.51, 2003.55 and 2003.58 (Figures \ref{fig:jan99} to \ref{fig:jul03_3}). The early observations of 1998 December provided the first detection of a magnetic field on \hd. Several images were reconstructed from subsets of the large data sets secured in summers 2002 and 2003, to take into account the short-term variability of the photosphere occurring during data collection (see a discussion of this point in Sect \ref{sect:changes}).

Owing to the relatively low \sn\ of the 2001.54 data set (3600 in average in Stokes V, i.e. 40\% below the mean \sn\ level of epoch 1999.02 for instance), none of the polarized profiles secured in this epoch provide a magnetic field detection, thus only a brightness image was constructed (Fig. \ref{fig:jul01}). All the surface topologies described hereafter are reconstructed with a reduced \kis (hereafter \kisr) of 0.65 and 0.9 for brightness and magnetic images respectively. A \kisr\ smaller than unity is adopted to take into account the fact that error bars derived from MuSiCoS observations are slightly overestimated (Wade et al. 2000).

\begin{figure*} 
\centerline{\psfig{file=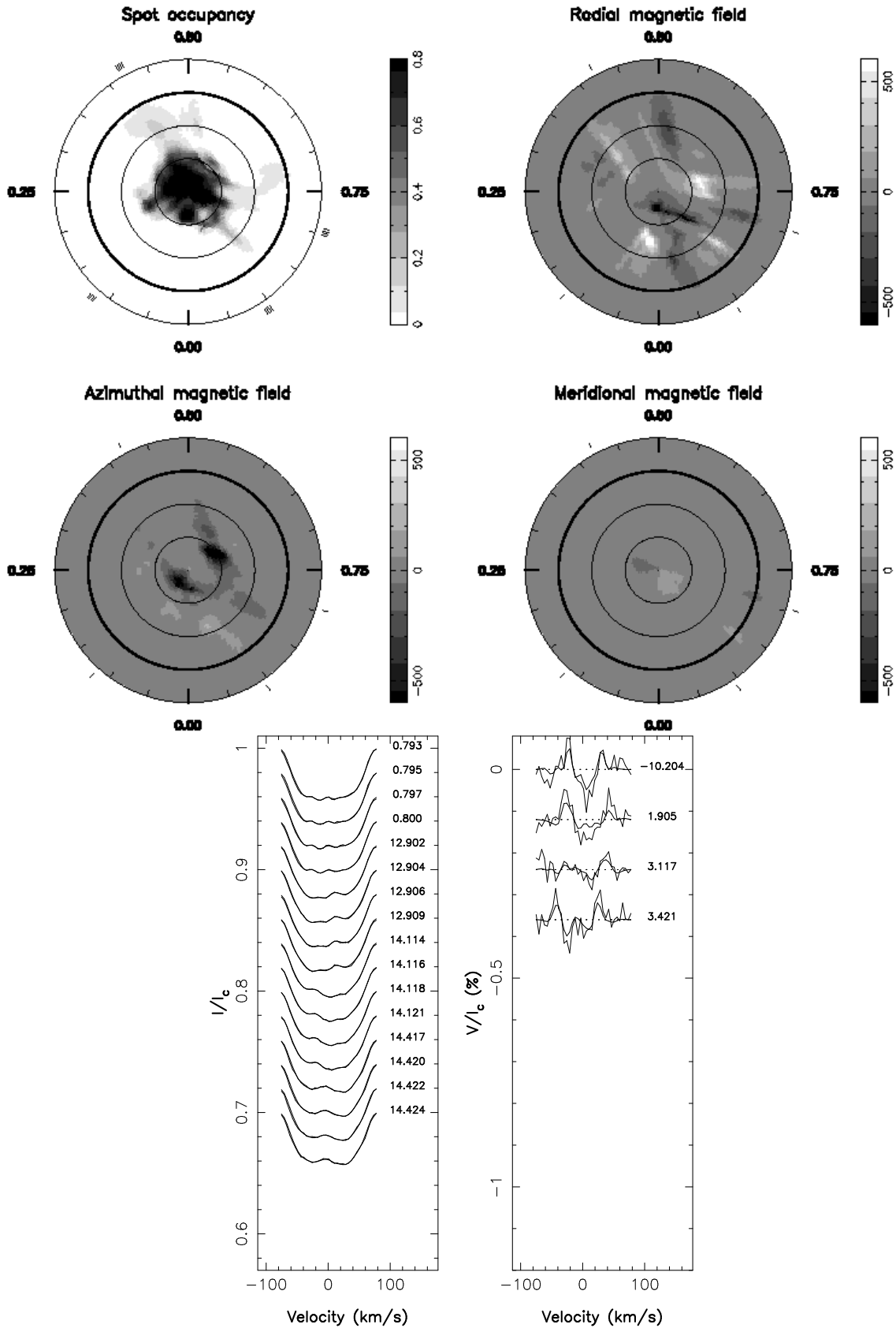,width=15cm}} 
\centerline{\mbox{\psfig{file=hd199178I_jan99.ps,height=10cm,angle=270} 
\hspace{2mm}      \psfig{file=hd199178V_jan99.ps,height=10cm,angle=270}}} 
\caption[]{Reconstructed images of HD~199178 at epoch 1999.02, in flattened polar view. The concentric circles correspond (starting from the outside) to parallels of latitude $-30$\degr, 0\degr (equator, bold line), $+30$\degr and $+60$\degr. The upper-left panel corresponds to a brightness image, while the three other charts show the components of the magnetic field (in Gauss) in spherical coordinates, i.e. radial, azimuthal and meridional components of the field in the upper-right, lower-left and lower-right charts respectively. Stokes I and V normalized profiles are plotted in the lower part of the figure (left-hand panel and right-hand panel, respectively). Thin lines represent the observed profiles, while bold lines correspond to profiles reconstructed by the imaging code.} 
\protect\label{fig:jan99} 
\end{figure*} 

\begin{figure} 
\centerline{\psfig{file=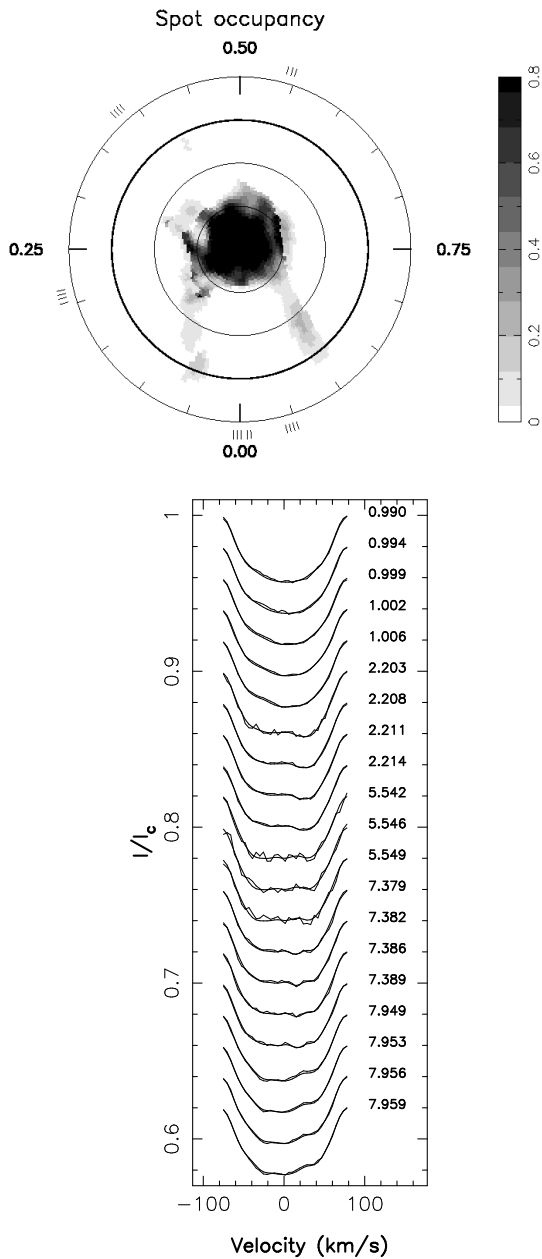,height=6cm} }
\vspace{5mm}\centerline{\psfig{file=hd199178I_jul01.ps,width=4cm,angle=270}} 
\caption[]{Same as Fig \ref{fig:jan99} for the 2001.54 data set. There was no Zeeman signature detected at this epoch, owing to a high noise level of Stokes V profiles (see Table \ref{tab:journal}).} 
\protect\label{fig:jul01} 
\end{figure} 

\begin{figure*} 
\centerline{\psfig{file=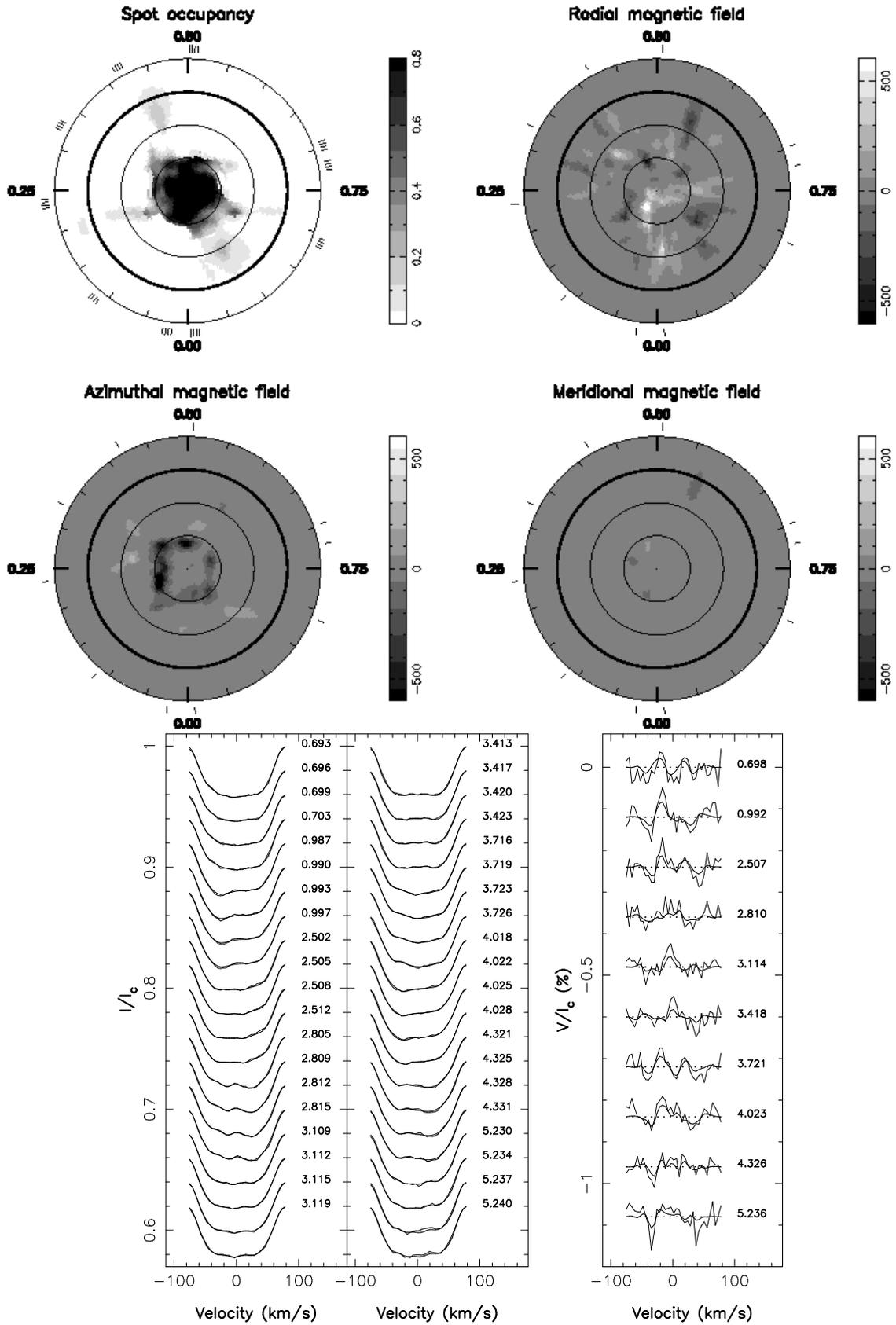,width=15cm}} 
\centerline{\mbox{\psfig{file=hd199178I_dec01.ps,height=10cm,angle=270} 
\hspace{2mm}      \psfig{file=hd199178V_dec01.ps,height=10cm,angle=270}}} 
\caption[]{Same as Fig \ref{fig:jan99} for the 2001.97 data set.} 
\protect\label{fig:dec01} 
\end{figure*} 

\begin{figure*} 
\centerline{\psfig{file=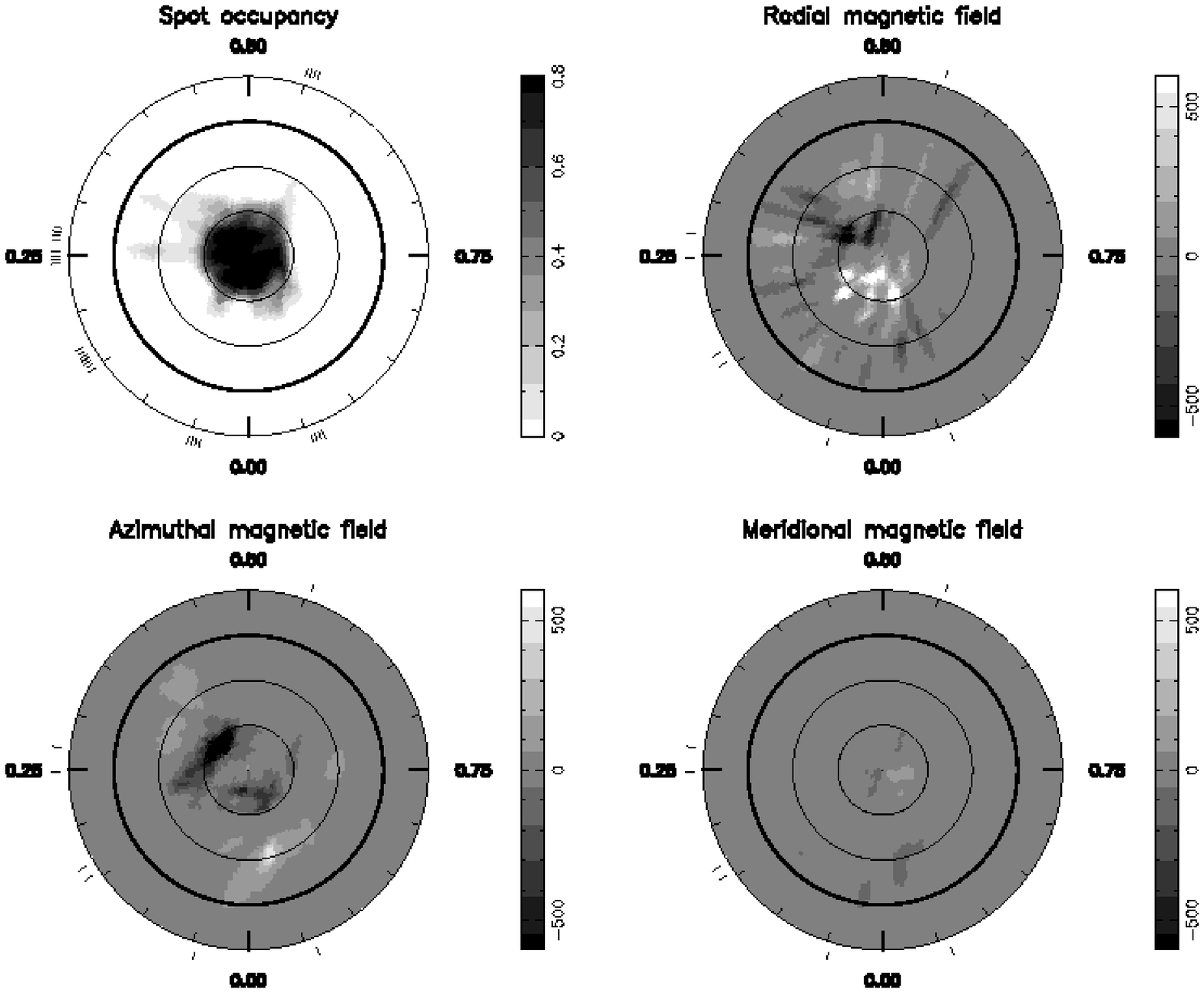,width=15cm}} 
\centerline{\mbox{\psfig{file=hd199178I_jun02.ps,height=10cm,angle=270} 
\hspace{2mm}      \psfig{file=hd199178V_jun02.ps,height=10cm,angle=270}}} 
\caption[]{Same as Fig \ref{fig:jan99} for the 2002.46 data set.} 
\protect\label{fig:jul02_1} 
\end{figure*} 

\begin{figure*} 
\centerline{\psfig{file=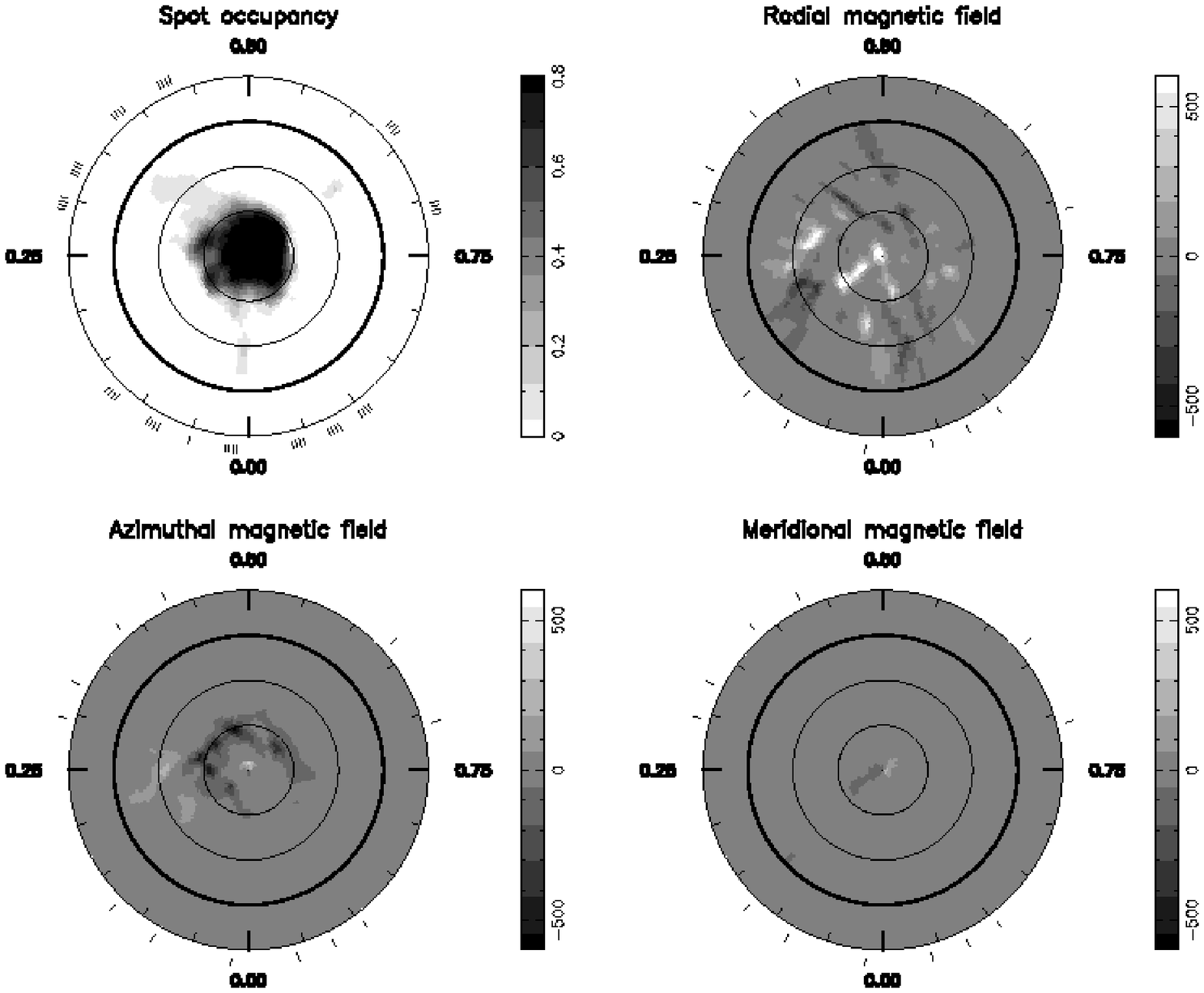,width=15cm}} 
\centerline{\mbox{\psfig{file=epoch1I.ps,height=10cm,angle=270} 
\hspace{2mm}      \psfig{file=epoch1V.ps,height=10cm,angle=270}}} 
\caption[]{Same as Fig \ref{fig:jan99} for the 2002.50 data set.} 
\protect\label{fig:jul02_2} 
\end{figure*} 

\begin{figure*} 
\centerline{\psfig{file=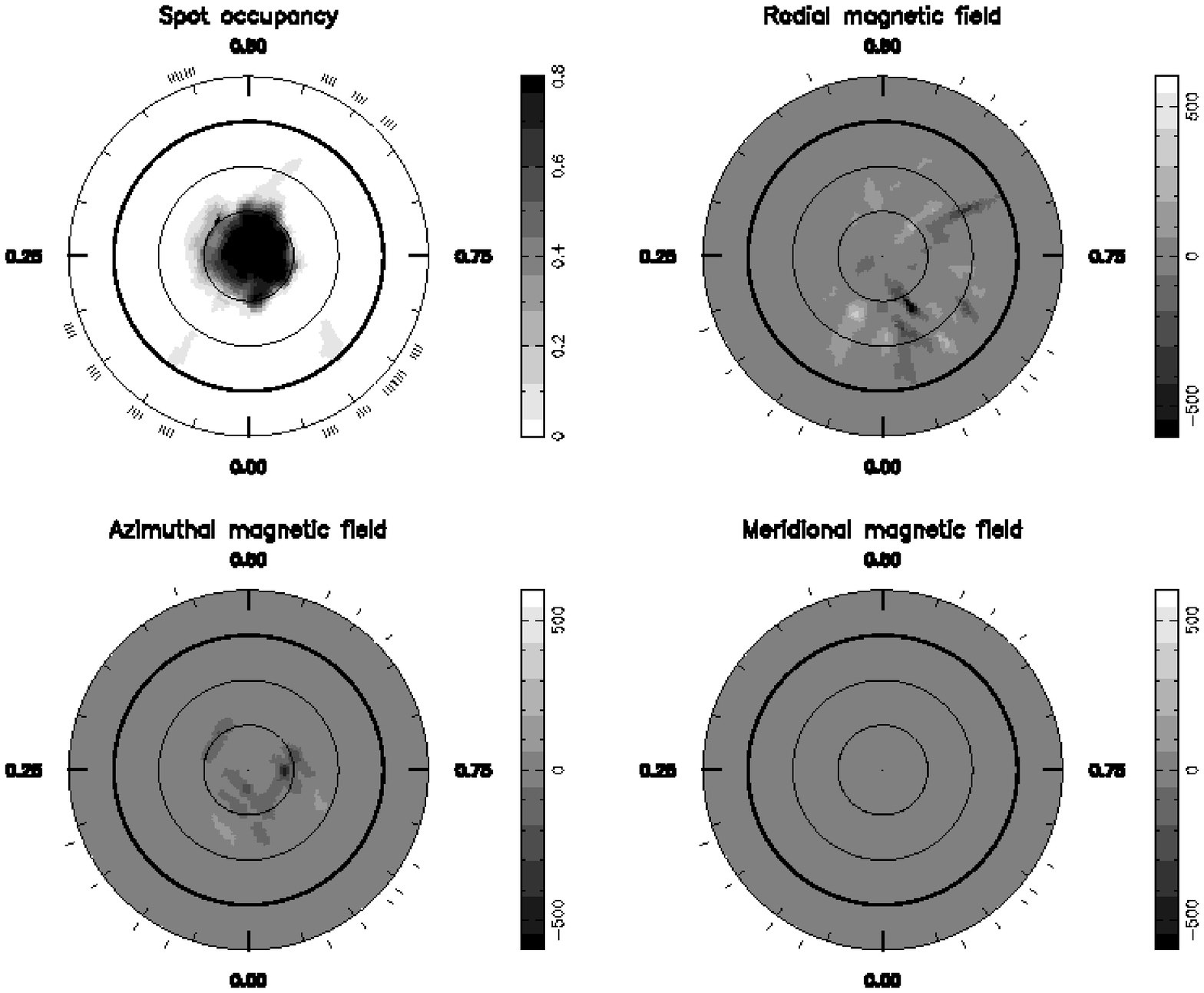,width=15cm}} 
\centerline{\mbox{\psfig{file=epoch2I.ps,height=10cm,angle=270} 
\hspace{2mm}      \psfig{file=epoch2V.ps,height=10cm,angle=270}}} 
\caption[]{Same as Fig \ref{fig:jan99} for the 2002.56 data set.} 
\protect\label{fig:jul02_3} 
\end{figure*} 

\begin{figure*} 
\centerline{\psfig{file=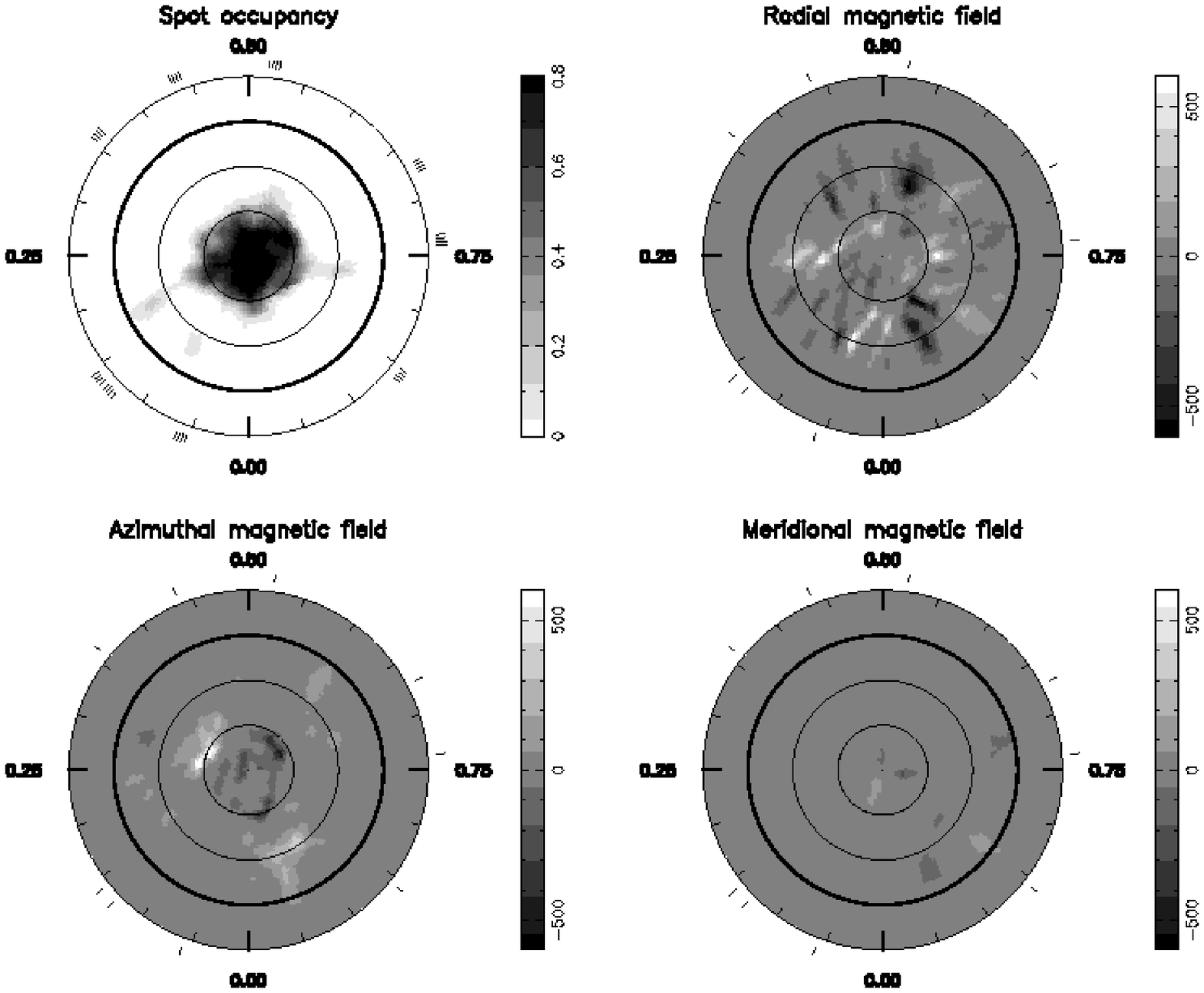,width=15cm}} 
\centerline{\mbox{\psfig{file=1jul03I.ps,height=10cm,angle=270} 
\hspace{2mm}      \psfig{file=1jul03V.ps,height=10cm,angle=270}}} 
\caption[]{Same as Fig \ref{fig:jan99} for the 2003.51 data set.} 
\protect\label{fig:jul03_1} 
\end{figure*} 

\begin{figure*} 
\centerline{\psfig{file=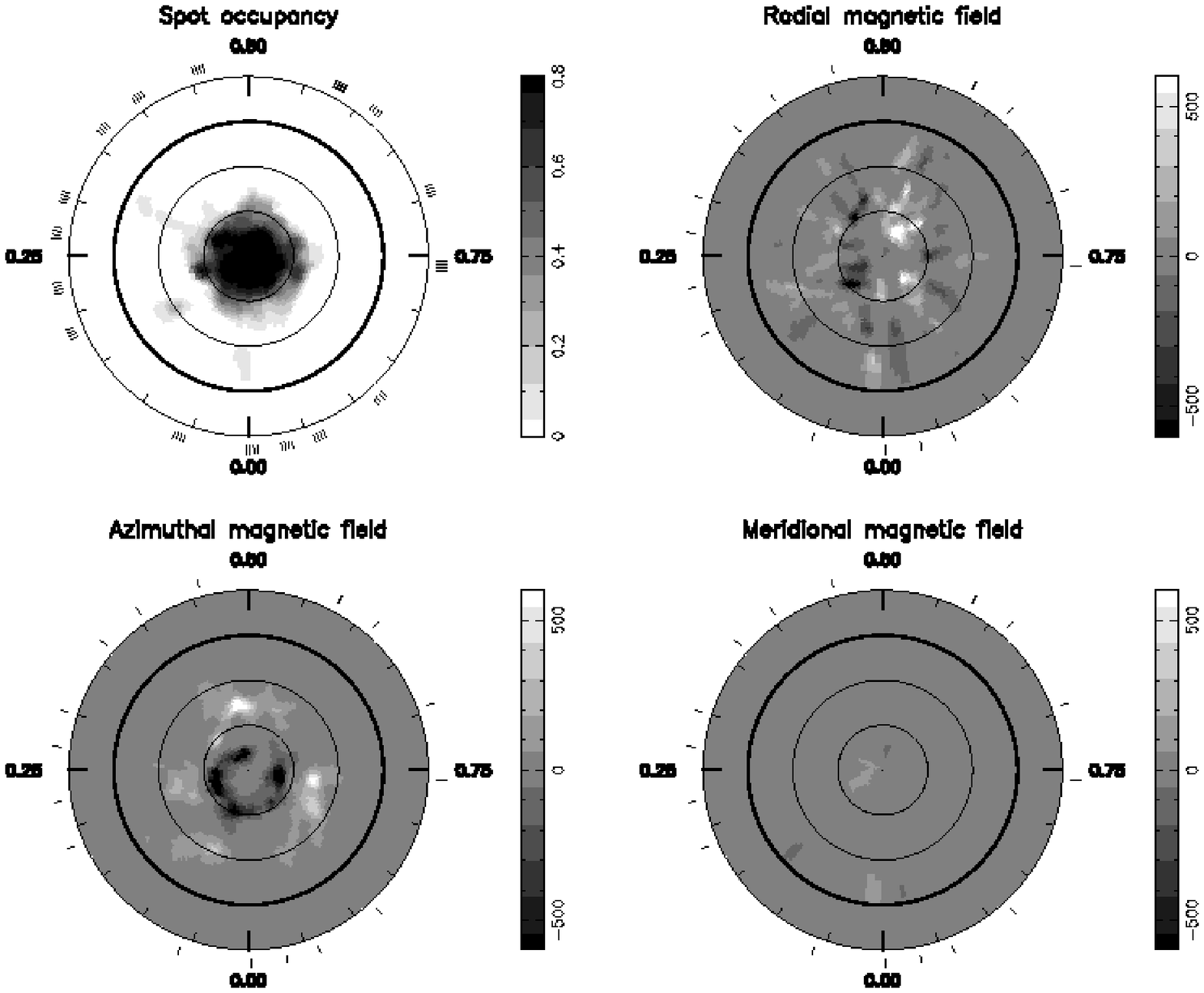,width=15cm}} 
\centerline{\mbox{\psfig{file=2jul03I.ps,height=10cm,angle=270} 
\hspace{2mm}      \psfig{file=2jul03V.ps,height=10cm,angle=270}}} 
\caption[]{Same as Fig \ref{fig:jan99} for the 2003.55 data set.} 
\protect\label{fig:jul03_2} 
\end{figure*} 

\begin{figure*} 
\centerline{\psfig{file=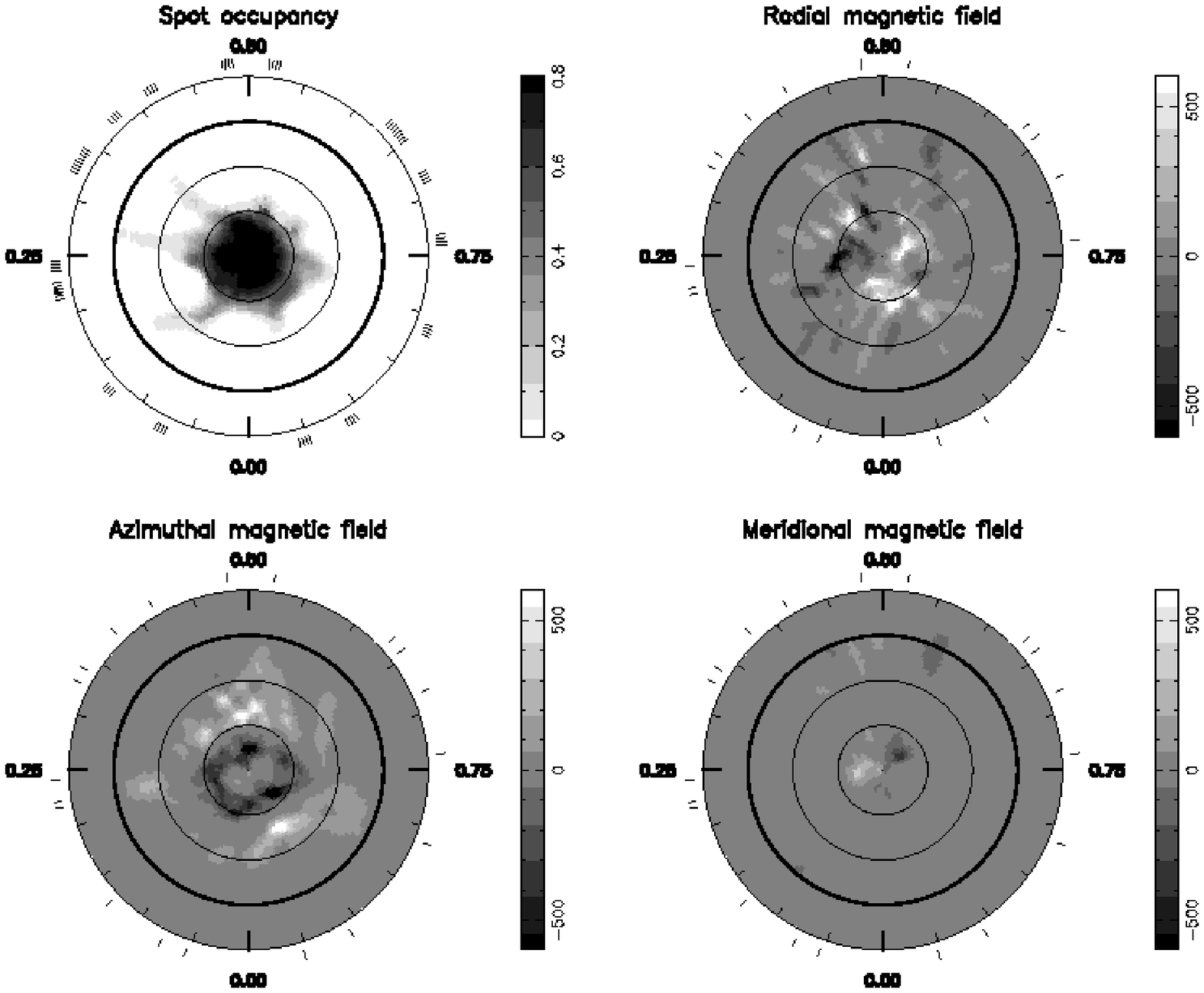,width=15cm}} 
\centerline{\mbox{\psfig{file=3jul03I.ps,height=10cm,angle=270} 
\hspace{2mm}      \psfig{file=3jul03V.ps,height=10cm,angle=270}}} 
\caption[]{Same as Fig \ref{fig:jan99} for the 2003.58 data set.} 
\protect\label{fig:jul03_3} 
\end{figure*} 

\subsection{Brightness surface structures}
\label{sect:brightness}

\begin{figure}  
\centerline{\psfig{file=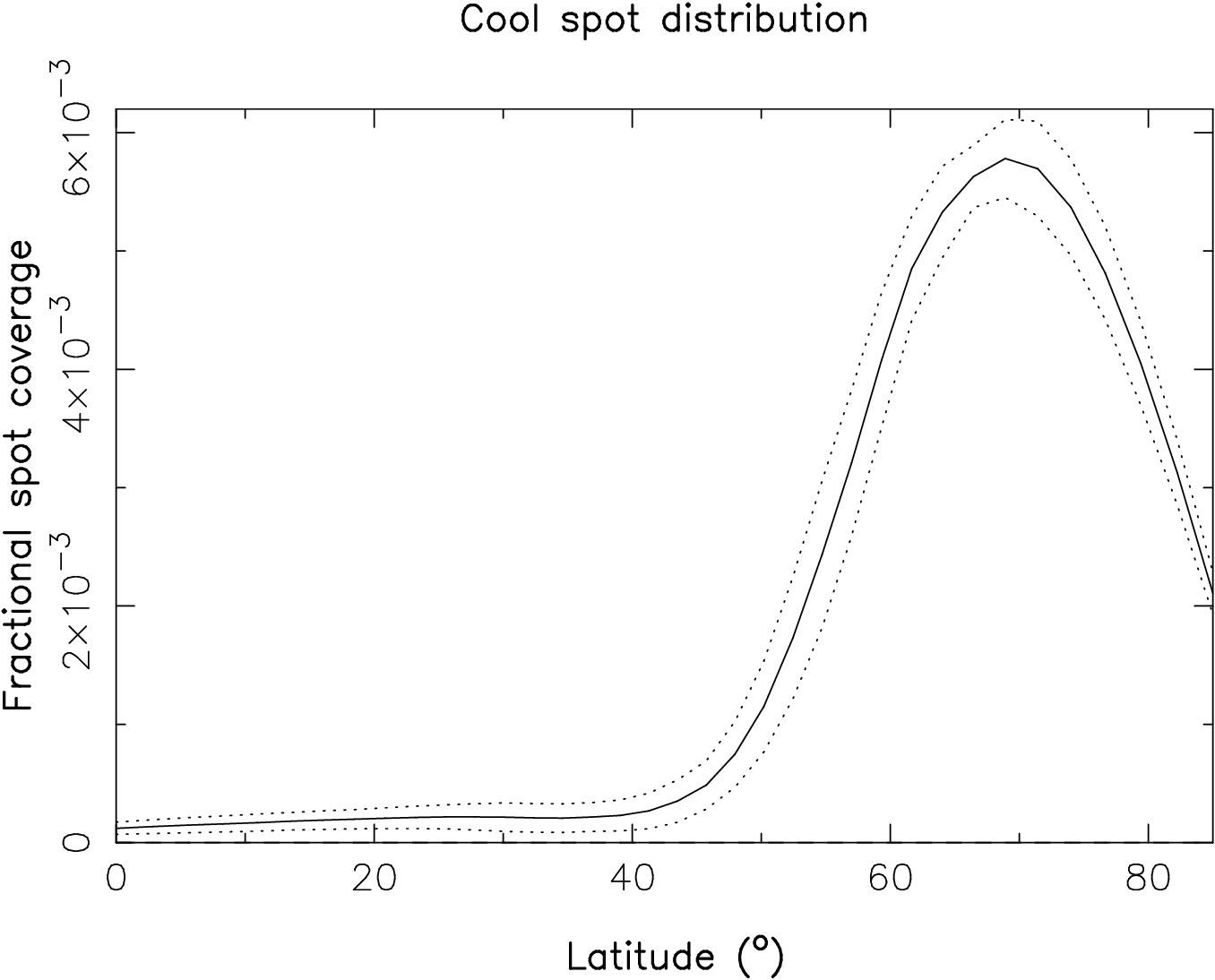,width=8cm,angle=270}}  
\caption[]{Distribution of the fractional spot coverage as a function of latitude. The full line curves represent the flux averaged over all observing periods. The two dotted curves surrounding the mean curve show the standard deviation.}  
\protect\label{fig:lat_spot}  
\end{figure}  

The most obvious characteristic of the brightness topology of \hd\ is the presence of a large polar spot in all images. This cool region is always quasi-axisymmetric and covers most of the stellar photosphere above latitude 60\degr. At several epochs, the polar spot is partly fragmented, with smaller spots clearly distinguishable from the main polar component (e.g. at epoch 1999.02, when a smaller, secondary spot was visible at phase 0.0 and latitude 80\degr).

In addition to the polar spot, smaller spots appear at lower latitude, but only about 5\% of the overall spot coverage is contained within the low latitude features (below latitude 30\degr). The latitudinal spot distribution is rather stable from one epoch to the next (see Fig. \ref{fig:lat_spot} for epoch 2002.50) and repeatedly presents a peak of spot occupancy between latitudes 15\degr\ and 30\degr, with a relative gap of spottedness in regions next to the equator and within a band extending from latitude 30\degr to the outer limit of the large polar cap. 

\begin{figure*} 
\centerline{\mbox{\psfig{file=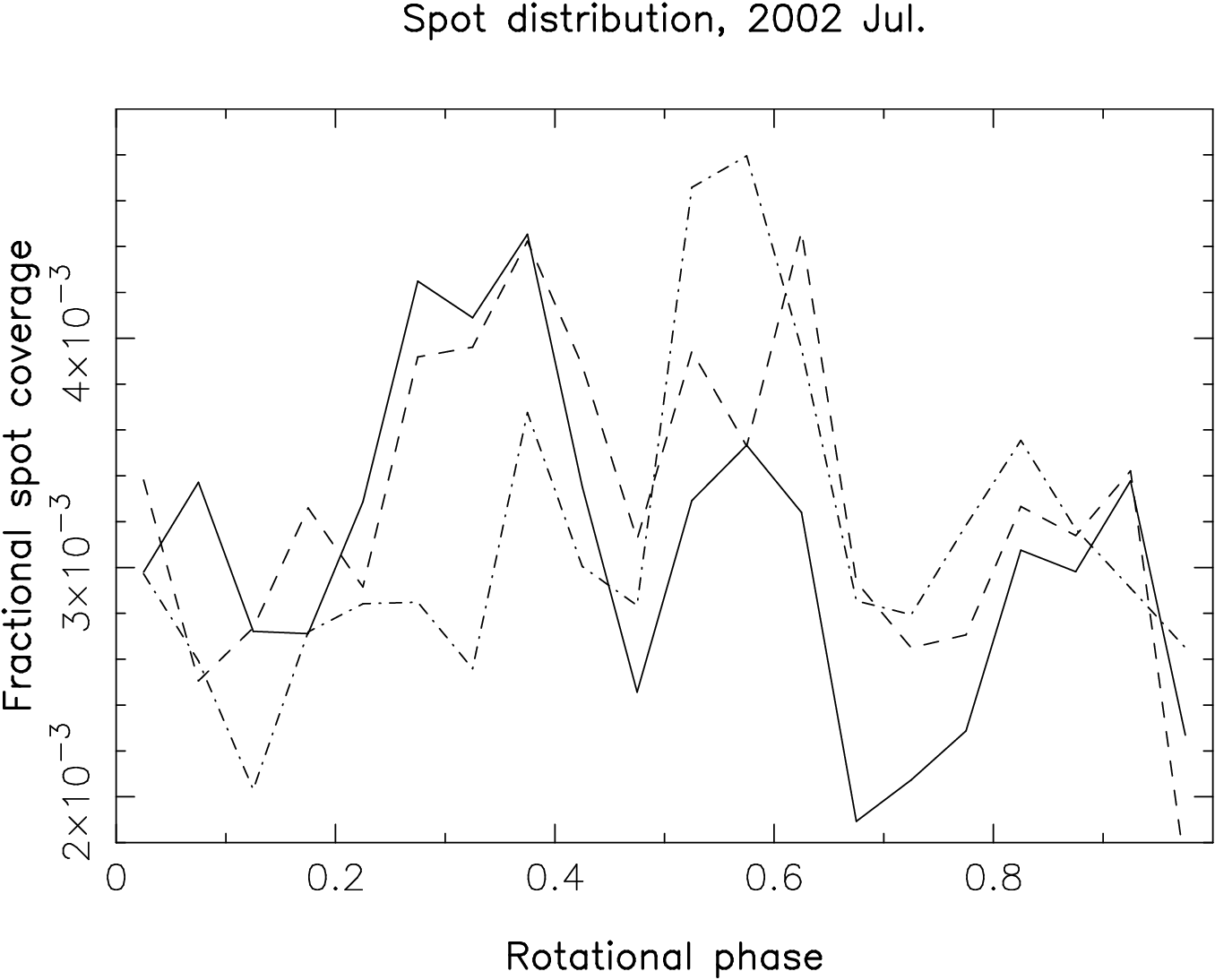,height=8cm,angle=270} 
\hspace{2mm}      \psfig{file=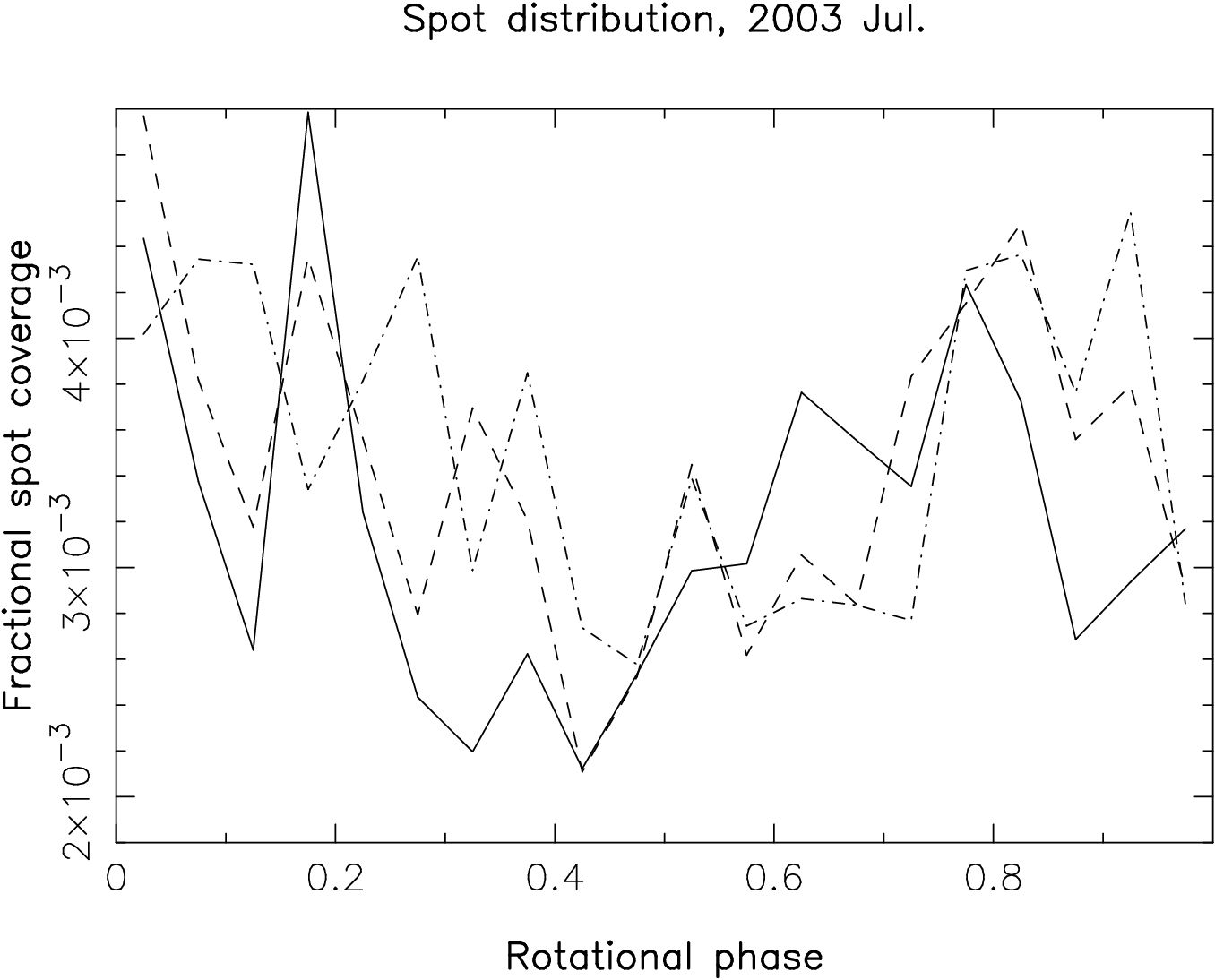,height=8cm,angle=270}}} 
\caption[]{Distribution of the fractional spot coverage as a function of the rotational phase in 2002 and 2003 (left-hand and right-hand panels, respectively). Individual curves are derived from successive data subsets (2002.46 and 2003.51 in full line, 2002.50 and 2003.55 in dashed line, 2002.56 and 2003.58 in dot-dashes).} 
\protect\label{fig:active_spot} 
\end{figure*} 

\begin{figure*} 
\centerline{\mbox{\psfig{file=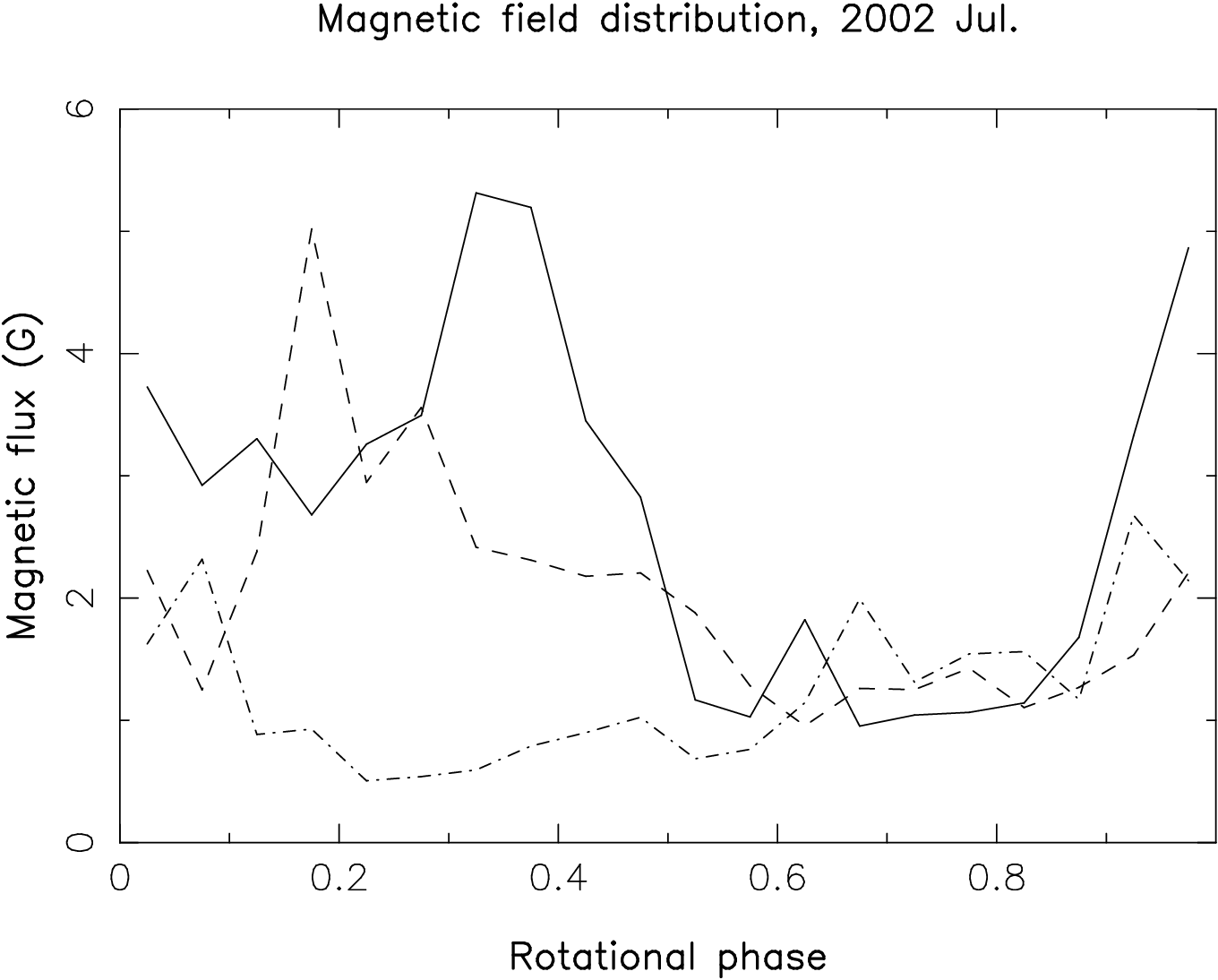,height=8cm,angle=270} 
\hspace{2mm}      \psfig{file=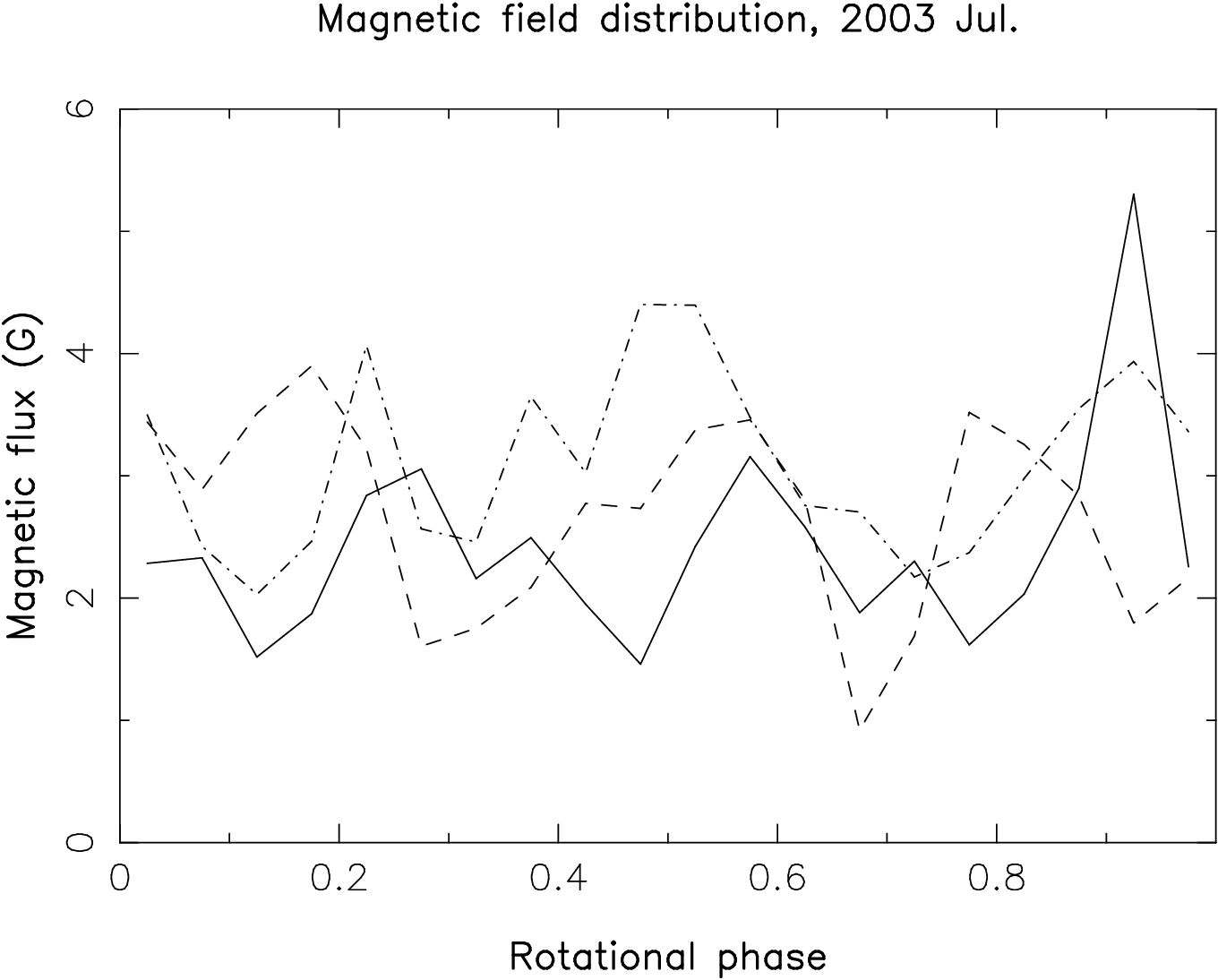,height=8cm,angle=270}}} 
\caption[]{Same as Fig \ref{fig:active_spot} for the fractional magnetic flux. } 
\protect\label{fig:active_mag} 
\end{figure*} 

In Fig. \ref{fig:active_spot} we plot the latitudinally-averaged spot coverage, as a function of the rotational phase, for each subset of summer 2002 (left panel) and summer 2003 (right panel). The first information provided by these curves is that the observed spot distribution is highly varying between near-in-time subsets. Part of this apparent evolution can be explained by the different phase sampling achieved for different subsets, producing a lack of signal at some longitudes. However, if phase gaps can be a problem at some epochs (e.g. 2002.46), most of the subsets possess good phase sampling, particularly in 2003. In order to obtain a more precise estimate of potential biases due to phase gaps, we compare successive subsets secured at close-by epochs with a different phase sampling (but bearing in mind however that short-term evolution of the spot distribution also accounts for a part of the observed differences, see Sect. \ref{sect:changes}). 

The spot coverage was marginally higher in 2002 on phases ranging from 0.25 to 0.65 (with two distinct maxima of spot coverage visible around phases 0.35 and 0.55). In 2003, the maximum spottedness is rather concentrated between phases 0.8 and 0.2. In 2002, longitudes with the largest filling factor are clearly linked to the presence of low-latitude spots (see Fig. \ref{fig:jul02_1} to \ref{fig:jul02_3}), while in 2003 features spatially associated to the large polar cap are mostly responsible for the higher spottedness around phase 0.0. For other epochs, the biases due to uneven phase coverage cannot be as accurately estimated as for the 2002 and 2003 data sets. However, if the phase sampling is far from optimal at epochs 1999.02 and 2001.54, there is no significant phase gap at epoch 2001.97, for which we note a maximum spot coverage around phase 0.9 (corresponding plot not shown here).

\subsection{Magnetic topology}
\label{sect:topology}

The magnetic topology of HD~199178 features several regions of intense field strength (in excess of 1~kG). Owing to the fact that polarized profiles used here suffer from a relatively high noise level (compared to previous studies of active stars based on ZDI), the constraints provided on the magnetic maps are not very high, so that several assumptions on the field are compatible with an image reconstruction at a fixed \kisr\ (equal to 0.9). For instance, the maps can be calculated assuming either a purely radial field or a mixture of all three possible components of the field (radial, azimuthal and meridional). To select between these two possibilities, the very principle of maximum entropy imaging tells us that the image exhibiting the lowest information content is the most likely (the information contained in the map being calculated as the integrated field strength over the stellar surface, thereafter B$_{\rm mod}$). Taking epoch 2003.58 as an example, a purely radial photospheric field leads to B$_{\rm mod}$=74~G. For the same data set, B$_{\rm mod}$ is equal to 46~G for an image containing all three components of the field. Very similar results can be derived for all other epochs, with B$_{\rm mod}$ about 30 to 60\% higher for a purely radial field than for the map containing all three field components. At this stage, it can be argued that assuming the presence of an azimuthal component is not absolutely necessary for computing the magnetic maps. However, several additional reasons lead us to consider this component as genuine, as detailed below and in Sect. \ref{sect:discuss}. 

The reconstructed azimuthal field component can itself be divided into a couple of subcomponents of opposite polarities, with a predominance of clockwise field at high latitude, the lower latitudes being dominated by a counter-clockwise magnetic field. The left panel of Fig. \ref{fig:field_dist} shows the longitudinally-integrated distribution of the azimuthal field (averaged over all observing epochs). Both sub-components can be clearly distinguished, with the field polarity changing sign around latitude 50\degr. The fact that the azimuthal component presents a consistent large-scale geometry between different images (despite very different \sn\ or phase coverage) suggests that the observed structures cannot be attributed to a problem of data quality (like noise or phase covering). 

The clockwise sub-component shows up in all images. At epoch 2001.97 for instance, a succession of several blobs of intense field strength draws the outline of a ring encircling the pole at the outer border of the cool polar spot. We also see an almost complete ring at epochs 2003.55 and 2003.58, whereas images reconstructed at other observing periods feature only a few spots of clockwise field, most of the time coincident with the edge of the polar spot. At some epochs (e.g. 2003.58), it can be noticed that spots belonging to the clockwise sub-component show up within the dark polar cap, whereas we expect very few polarization signal from this region because of the important brightness contrast with the unspotted photosphere (at a fixed field strength, Zeeman signatures formed inside the polar cap will be much fainter than that formed in the surrounding photosphere). Some of these structures may indeed belong to the outer region of the polar spot, but reconstruction biases of the imaging code can slightly shift these magnetic regions toward higher latitudes (Petit et al. 2002). It cannot be excluded however that Zeeman signatures are observed inside the polar spot. In this case, the reconstructed magnetic field must be considered with a caution, since the field strength is likely underestimated. 

The counter-clockwise component of the azimuthal magnetic field appears more discreetly between latitudes 10\degr\ and 40\degr. At epoch 2003.58 for instance, several spots of azimuthal field appear around latitude 40\degr, but with a relatively low field strength (reaching about 0.6~kG at phases 0.5 and 0.9). The low level of the magnetic flux emerging from these regions (Fig. \ref{fig:field_dist}) explains why such components cannot (or only marginally) be reconstructed at epochs for which the data suffer from a lower \sn\ (epochs 2002.50 and 2002.56 for instance). 

In addition to areas hosting an azimuthal magnetic field, we reconstruct several regions in which the field is mostly radially oriented. On most images, the radial field is organized in an intricate pattern mixing opposite polarities, with a large-scale structure much less obvious than that of the azimuthal component. We note also that the radial field regions are varying a lot between images corresponding to close-by epochs, which may lead to question their reality. Numerical simulations suggest indeed that the noise pattern is preferentially reconstructed as radial field regions (but this effect is expected to be significant only for \sn\ much lower than that achieved here). As a test, we divide all the epoch 2003.58 data into two independent data sets by taking the odd numbered spectra to generate one image and the even numbered spectra to generate a second image (both images not shown here). The smallest radial field regions obtained in the corresponding images show little similarities between both maps, while all larger structures are coherently reconstructed. We emphasize however that reconstructing a map with one half of the profiles is tantamount to increasing the noise by a factor $\sqrt 2$. Given the fact that the reconstructed magnetic distribution is therefore less accurate, we caution that discrepancies at small scales in the radial field pattern do not demonstrate that small-scale structures are not reliable in images better constrained. Furthermore, the good consistency achieved for all large structures is a good indicator of their reality. 

The largest radial field regions are observed at epochs 2002.46 and 2003.58, when two large unipolar areas are seen close to the pole. Contrary to the azimuthal field, no latitudinal organization of the radial field can be readily seen on the different images. However, if we average the radial field component over all observing epochs (Fig. \ref{fig:field_dist}), we note that a positive field dominates at high latitude, while regions presenting a negative polarity are mostly confined below latitude 40\degr. Despite this statistical trend, it can be seen on some images that a negative polarity can sometimes dominate the radial component at high latitude (at epochs 1999.02 and 2002.56 for instance).

A small part of the polar magnetic field is sometimes reconstructed as a meridional field component (see e.g. Fig \ref{fig:jul02_2}). These high latitude azimuthal and meridional features indeed belong to the same physical structure (a region containing mostly horizontal field and located over the pole), an effect (detailed by Donati 1999) that can be attributed to the singular nature of the pole in spherical coordinates. Other structures appearing in the meridional field component at lower latitude (see, e.g., very faint structures around phase 0.5 at epoch 2003.58) can be partly produced by known reconstruction cross-talks between radial and meridional fields (Donati \& Brown 1997), an effect essentially negligible in the case of \hd\ thanks to its relatively high inclination angle. 

The radial magnetic field is always the main contributor to the photospheric magnetic energy (at roughly 85\% of the overall energy). However, numerical simulations suggest that this fraction is sensitive to the \sn\ and may become high for low-quality data (part of the noise producing radial field structures, even for very slight over-fitting of the data). In order to avoid this problem, a solution consists in calculating the energy stored in the {\em longitudinally-integrated} field. By doing so, we average out most of the noise, but we also ignore most of the energy contained in small-scale magnetic regions. It is however an opportunity to estimate the energy stored within the large-scale axisymmetric field, which is otherwise partly hidden among small magnetic features. We find that the radial field contributes only to some 10\% of the global energy contained in the large-scale field (at epochs 2001.97 and following), which means in other words that the axi-symmetric field is largely dominated by its azimuthal component.

At epochs 2002.46, 2002.50 and 2002.56, the low-latitude ring of azimuthal field steadily contains about 14\% of the total amount of energy stored within the azimuthal component of the axisymmetric field. This fraction is stable despite large variations of the \sn\ and phase coverage over the 3 corresponding data sets, suggesting that data quality only marginally impacts this estimate. The fraction was equal to 6\% six months earlier at epoch 2001.97 (with significantly better \sn\ and comparable level of phase coverage) and reaches about 20\% between 2003.51 and 2003.58. We suggest that this evolution may be genuine, with respective weights of both azimuthal field rings remaining rather stable over a few weeks, but showing significant evolution on timescales ranging from a few months to several years. 

The latitudinally-averaged distribution of the photospheric field as a function of the rotational phase is plotted in Fig. \ref{fig:active_mag} for data sets secured during summers 2002 and 2003. A larger magnetic flux is observed between phases 0.2 and 0.4 during summer 2002, but with significant differences between the three curves corresponding to epochs 2002.46, 2002.50 and 2002.56. We will discuss in Sect. \ref{sect:changes} whether short-term evolution of active regions can account for this apparent discrepancy. At epoch 2002.56 and during summer 2003, the distribution of the magnetic flux is mostly independent of the rotational phase. In particular there is no apparent correlation between the phase distributions of cool spots (Fig. \ref{fig:active_spot}) and magnetic field. 

\subsection{Rapid surface changes}
\label{sect:changes}

\begin{figure}  
\centerline{\psfig{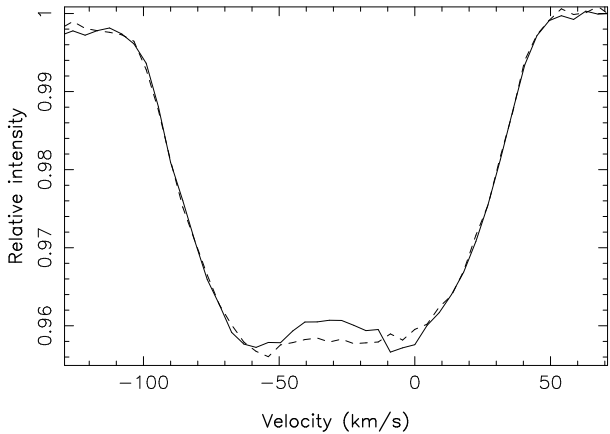}}  
\caption[]{Stokes I profiles of \hd\ obtained at phase 0.5588 on 2002 June 13 (dashed line), and at phase 0.5629 on July 26 (full line).}  
\protect\label{fig:2profiles}  
\end{figure}  

During summers 2002 and 2003, \hd\ was observed every clear night, thus providing a monitoring (over two distinct periods covering about 45~d) of the evolution of surface structures.

The intrinsic evolution of the photosphere is first obvious in the profiles themselves. For instance, striking differences are visible between the brightness profiles secured on 2002 June 13 and 2002 July 26, i.e. 43 days apart (Fig. \ref{fig:2profiles}), despite a difference in rotational phase as small as 0.4\% of a rotation cycle. In this specific case, the wide bump appearing at the center of the most recent profile has no equivalent 6 weeks before. 

The second evidence for surface variability arises from a monitoring of the \kisr\ of the reconstructed images. When grouping the data secured from 2002 June 28 to 2002 July 28 in a single data set (i.e. grouping epochs 2002.50 and 2002.56), we cannot reconstruct surface images with a \kisr\ lower than 0.85 and 1 for the brightness and magnetic images respectively, as opposed to a \kisr\ of 0.65 and 0.9 when reconstructing separate images from each sub-set. To ensure that the data sets employed to produce the surface images do not spread on too long a timespan, we split the global data sets of summers 2002 and 2003 into successive subsets corresponding to epochs 2002.46, 2002.50, 2002.56, 2003.51, 2003.55 and 2003.58 (respectively covering 14, 13, 11, 15, 13 and 11 nights). We then use this time series to get a direct view at the surface evolution of brightness and magnetic structures. In the following paragraphs, we voluntarily limit the comparisons between pairs of images to high-latitude structures and to the low-latitude regions located at rotational phases observed in both data sets, in an attempt to avoid as much as possible differences produced by reconstruction biases associated to phase gaps.

Considering the set of brightness images, we first note the continuous birth and disappearance of low latitude spots. For instance, the spot located at phase 0.4 and latitude 70\degr\ at epoch 2003.58 cannot be associated to any structure observed at epochs 2003.51 and 2003.55. On the other hand, the spot reconstructed around phase 0.17 and latitude 25\degr\ at epochs 2003.51 remains clearly visible until 2003.58 (small changes in its exact location and shape from one epoch to the next being compatible with usual reconstruction biases).

Rapid changes also affect brightness inhomogeneities at the limit of the polar spot during similar timescales. The edges of the polar cap (from phases 0.0 to 0.5) are rapidly evolving during the summer of 2002, with two secondary spots appearing next to the main component of the polar spot between epochs 2002.46 and 2002.50, then followed by a third region, located around phase 0.98 at epoch 2002.56. Note that these small spots are located around latitude 60\degr\ and therefore are sufficiently far from the pole to produce significant rotational modulation in the spectra. Considering also that these structures are never eclipsed during stellar rotation, we can conclude that the location of such cool spots is very well constrained, so that the evolution we report here is very likely to be real.

The same type of evolution is observed for magnetic regions, with striking changes in the magnetic topology occurring on timescales as short as a couple of weeks. One of the most obvious change concerns the large azimuthal field region located at phase 0.5 and latitude 45\degr\ at epochs 2003.55 and 2003.58, which has no counterpart at epoch 2003.51. Another striking example concerns the very intense, large spot located at phase 0.4 and latitude 60\degr\ at epoch 2002.46. The field strength inside this region, in excess of 1~kG at epoch 2002.46, rapidly decreases to about 100~G in epoch 2002.56 (this evolution explains part of the differences between curves plotted in the left panel of Fig. \ref{fig:active_mag}). We caution however that the relatively low \sn\ of data sets corresponding to epochs 2002.50 and 2002.56 may also be partly responsible for the observed evolution. 

Concerning the spots of radially-oriented magnetic field, only the largest structures seem to possess lifetimes larger than a couple of weeks. As an illustration, the large region of positive field located near the pole at epoch 2002.46, which has already partly disappeared at epoch 2002.50, is then replaced by a region of opposite polarity at epoch 2002.56. We must nevertheless keep in mind that the location and field strength of magnetic regions inside the dark polar spot are likely to be affected by reconstructions biases, as already outlined in Sect. \ref{sect:topology}.

\begin{figure*}
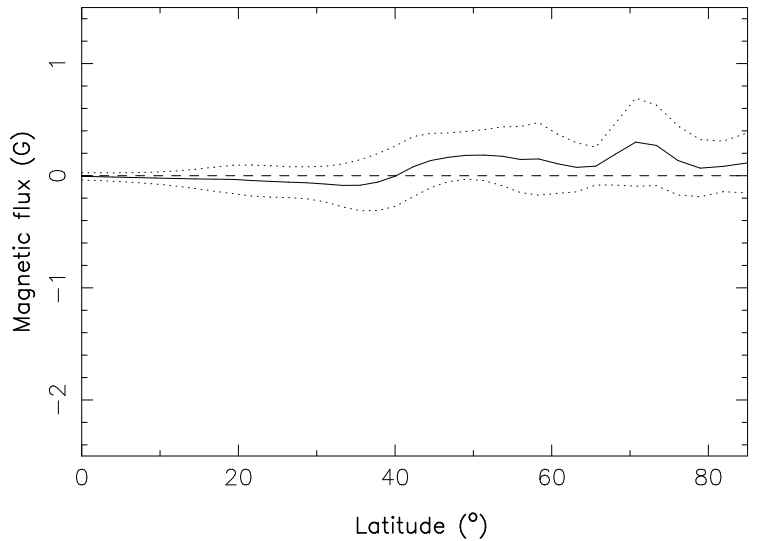
  
\centerline{\mbox{\psfig{file=azim.ps,height=8cm,angle=270} 
\hspace{2mm}      \psfig{file=radial.ps,height=8cm,angle=270}}} 
\caption[]{Latitudinal distribution of the azimuthal (left) and radial (right) components of \hd\ magnetic flux. The full line curves represent the flux averaged over all observing periods. The dotted curves show the standard deviation.}  
\protect\label{fig:field_dist}  
\end{figure*}

\section{Differential rotation} 
\label{sect:diffrot} 

In this section, we study the motion of surface structures under the effect of differential rotation, applying the dedicated method of Petit et al. (2002) and assuming a surface rotation law similar to that given by equation \ref{eq:diffrot}. Considering a time-series of profiles, up to 2,000 images are computed (either brightness or magnetic maps, depending whether polarized or unpolarized profiles are considered) assuming different values of the differential rotation parameters \omeq\ and \dom\ and imposing a constant information content in all images. In the case where the data quality is good enough, the reconstructed images present a \kisr\ minimum corresponding to the most likely set of \drot\ parameters. Considering a \kisr\ map obtained from the whole set of computed images (e.g. Fig. \ref{fig:chi2map}), the region around the \kisr\ minimum can be approximated by a paraboloid, which curvature radius around its minimum gives the formal error bars on both parameters \omeq\ and \dom\ (Donati et al. 2003b). 

\begin{table}
\caption[]{Surface differential rotation parameters derived for \hd, using Stokes V profiles. For each epoch, \omeq\ (equatorial rotation rate) and \dom\ (difference of rotation rate between equator and pole) are listed.}
\begin{tabular}{ccc}
\multicolumn{1}{c}{Date}  &   \omeq   &   \dom \\
year  &   \rpd    &  \mrpd \\
\hline
2001.97 & $1.934 \pm 0.018$ & $66 \pm 28$ \\
2002.48 & $1.932 \pm 0.033$ & $79 \pm 61$ \\
2003.57 & $1.931 \pm 0.017$ & $86 \pm 20$\\
\hline
\end{tabular}
\label{tab:diffrot}
\end{table}

Only three of our data sets provided a detection of differential rotation (Table \ref{tab:diffrot}). The first one corresponds to observations obtained at epoch 2001.97. The two others are subsets taken from the large data sets of summers 2002 and 2003. In summer 2002, we consider profiles secured between June 15 and July 2. In summer 2003, the subset is built from observations ranging from July 21 to August 5. Subsets of 2002 and 2003 observations are chosen to present the best compromise between phase sampling, noise level and time-span. In particular, we ensure that the time-length of all data sets used to derive the \drot\ parameters is consistent with the typical timescale of local changes in the photospheric structures, as discussed in Sect. \ref{sect:changes}. Measurements of the \dom\ parameter give similar results in 2001.97 and 2003.57, with \dom\ equal to $66 \pm 28$~\mrpd\ and $86 \pm 20$~\mrpd\ respectively, indicating the detection of a solar-like \drot\ (the equator rotating faster than the pole) to within 4.3$\sigma$. The third estimate, corresponding to epoch 2002.48, suffers from error bars 2 to 3 times larger, but stays consistent with measurements from the two other epochs. 

No other data set constituted of Stokes V profiles could provide a detection of differential rotation (no paraboloid shape of the \kisr\ map around \kisr\ minimum). A higher noise level, a sparser phase sampling or a local evolution of the distribution of magnetic regions during data collection may be responsible for this failure. We also note that no detection was achieved using Stokes I profiles. This is not a surprise, since cool spots are mostly concentrated close to the pole on \hd, few tracers being available at lower latitude. On the contrary, the magnetic topology offers tracers densely distributed over the stellar surface, therefore spanning a larger range of rotation periods and allowing an easier detection of a surface shear. 

As outlined by Petit et al. (2002 and 2003), measurements of differential rotation can be partly affected by biases arising from low data quality. Also problematic is the potential aliasing due to intrinsic evolution of the photosphere (a newborn active region can be mistaken for an older one that has vanished during data collection, leading to inadequate evaluations of the shear). However, it is highly unlikely that an aliasing problem occurs for the three data sets presented here, and even more unlikely that the produced bias could shift our estimate of \drot\ toward the same (spurious) value. Given the fact that the data sets also present very different \sn\ and phase sampling, the three consistent measurements presented here suggest that the observed shear is genuine. 

\begin{figure}  
\centerline{\psfig{file=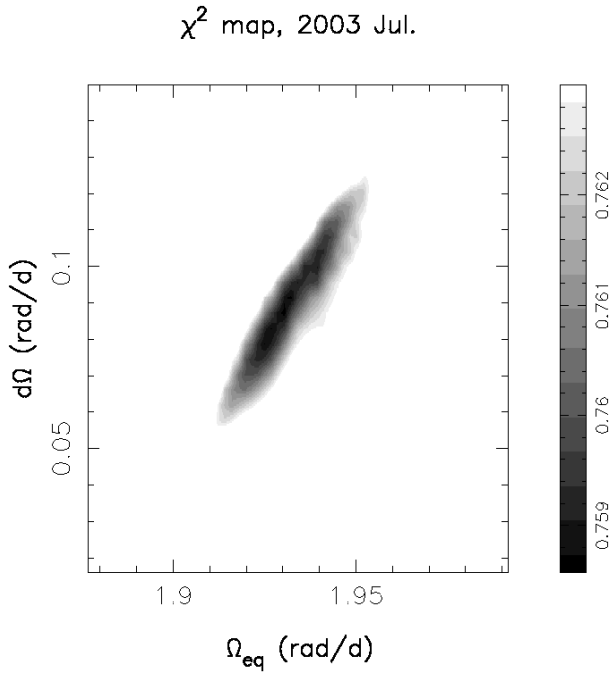,width=8cm,angle=270}}  
\caption[]{reduced $\chi^2$ map in the differential rotation parameter plane, obtained from polarized data of epoch 2003.57. $\Omega_{\rm eq}$ is the rotation rate of the equator and $\rm d\Omega$ the difference in rotation rate between the pole and the equator. The 1$\sigma$ limit on the two parameters (considered separately) corresponds to the black region.}  
\protect\label{fig:chi2map}  
\end{figure}  

\section{Discussion} 
\label{sect:discuss} 

\subsection{Brightness distribution}
\label{spots}

The magnetic activity of \hd\ has been monitored by the present study over 4.5~yr. Several characteristics of its brightness topology remain constant during this whole period, the most obvious being a large axisymmetric polar spot reconstructed above latitude 60\degr. While dark polar caps are known to be a usual feature on Doppler images of fast rotators, their exact shape and evolution with time can significantly vary for different objects. The only other evolved star for which ZDI has been performed up to now (the K1 sub-giant of the RS~CVn system \hr, see e.g. Petit et al. 2003) also possesses a cool polar region, but with a more complex shape and a fluctuating location of its centroid. Taking into account the regular, almost axi-symmetric shape of the polar spot observed on \hd, Hackman et al. (2001) argue that the flat bottom of the photospheric Stokes I profiles (interpreted by imaging codes as a polar spot) may indeed be partly due to a distortion of line profiles under the effect of anti-solar differential rotation (the polar region rotating faster than the equator). Our own measurements of the surface shear of \hd\ suggest a solar-like \drot\ of magnetic structures, arguing against this possibility. Moreover, numerical simulations show that for shear intensities similar to that we measure, line profile distortions do not exceed $10^{-4}$ of the continuum level, allowing us to safely assume that the \drot\ of \hd\ does not produce significant biases in the imaging process. 

\subsection{Magnetic topology}
\label{dynamo}

The magnetic topology of \hd\ presents at all observing epochs large regions in which field lines are mostly azimuthally oriented. Such magnetic regions have already been observed in several objects (e.g. Donati et al. 2003a) with sufficiently high \sn\ to erase any doubt on their reality. Here, two long-lived rings displaying opposite polarities are observed, encircling the pole at two distinct latitudes. The fact we observe this large-scale structure on all maps suggests that it cannot be attributed to an artifact due to the noise pattern. This general configuration of the azimuthal field is very similar to that of \hr, whereas no such obvious axi-symmetric structure has been reported for younger stars (which suggests that such large-scale characteristics of the magnetic topology are not a systematic bias produced by ZDI). The radial field component does not present a similarly axi-symmetric structure.


This predominant azimuthal component may be connected to the toroidal component of the large-scale dynamo field, by analogy with dynamo models developed for the Sun. However, this would imply at the same time that the dynamo operating in \hd\ is possibly different from that at work in the Sun. In the solar case, the toroidal field is believed to be mostly confined at the interface between the radiative core and the convective zone. The observation of this toroidal component at the photospheric level would therefore suggest that the dynamo of \hd\ is active either very close to its surface, or maybe in its whole convective envelope. Steep sub-surface velocity gradients are observed in the Sun (Corbard \& Thompson 2002) and it was investigated by Dikpati et al. (2002) whether a near-surface dynamo could be generated in this layer. They concluded however that such a mechanism could only marginally contribute to create a large-scale field. Recent simulations of K\"uker \& Stix (2001) however suggest that sub-surface velocity gradients may be stronger for stars with deep convective zones and for fast rotators. In this context, the possibility that a sub-surface dynamo may be efficient in a star like \hd\ is an interesting option that could partly account for the observed strong azimuthal photospheric field.

\subsection{Active longitudes}

Recent studies (based on long-term photometric observations) report the presence of active longitudes on FK~Com, a star very similar to \hd\ (Korhonen et al. 2002). Furthermore, the same authors report occasional 180\degr\ shifts of the activity peak longitude, a phenomenon usually called ``flip-flop''. The kind of monitoring we propose in the present study is not adapted to analyze such an effect, due to an inadequate time interval between successive images (a regular monitoring of \hd\ over several months would be an appropriate observational basis in this aim). In our observations there is only a marginal evidence that some rotational phases are more spotted than others (Fig. \ref{fig:active_spot}). It is also worth noticing that we do not detect any correlation between the phase distributions of spot coverage and magnetic flux (Fig. \ref{fig:active_mag}). This may suggest that the observed photospheric magnetic field is not directly connected to the internal field producing the cool spot pattern. In particular, it gives further support to the idea that the observed magnetic structures may be formed close to the stellar surface, while cool spots may reveal the action of a dynamo seated deeper in the stellar interior.

\subsection{Magnetic cycle}

The observed topology of the magnetic field is very stable over the whole observing period. While local modifications of the magnetic pattern occur on timescales as short as a couple of weeks, some general characteristics of the large-scale field (like the rings of azimuthal field) only undergo marginal changes on the 4.5~yr observing window. Only two noticeable long-term changes can be reported here. 

The first one is the progressive polarity reversal of the radial component of the field at high latitude, between epochs 2002.46 and 2002.56. No simultaneous variability of other large magnetic structures was noticed. A similar phenomenon was earlier reported for AB~Dor (Donati et al. 2003a) but over a longer timescale, of order of 1~yr. Even if we cannot exclude at this stage that this local evolution is part of the global activity cycle of \hd, the overall stability of the largest magnetic structures rather suggests that this kind of event is only local, thus not necessarily connected to a global variability of the large-scale field. The second noticeable evolution, which shows up on a timescale of several years, is the increasing fraction of magnetic energy stored at low latitude in the azimuthal field component.

\subsection{Surface \drot}
\label{discussdrot}

We report the probable detection of a solar-like \drot\ on \hd, derived by a monitoring of the relative motion of surface magnetic tracers. The intensity of the surface shear, estimated from the difference in rotation rate between equatorial and polar regions, is roughly solar in magnitude, with a lap-time (time for the equator to lap the polar region by one complete cycle) equal to 80~days. The repeated detection of \drot\ from independent data sets and the size of the related error bars allows us to rule out the possibility of an anti-solar \drot. 

The fact that the shear intensity is similar in magnitude to that measured on several fast rotating young dwarfs (Cameron et al. 2000) may indicate that a difference in the depth of the convective zone has only a limited impact on the shear level. It was also suggested by Cameron et al. (2000), considering a small sample of G and K active stars, that the lap-time may decrease for increasing stellar masses. With a mass slightly higher than that of other stars of this sample (a stellar mass of $1.65 M_{\odot}$ was proposed by Ackman et al. 2001), the lap-time of \hd\ is also among the shortest measured on other active G and K stars, therefore giving further support to this idea. 

A very weak solar-like surface shear similar to that detected on \hr\ (Petit et al. 2003 report for this star a lap-time equal to 480~d) is excluded to the $4\sigma$ level. Both \hr\ and \hd\ are evolved stars with deep convective envelopes, therefore we do not expect a difference in their evolutionary stage to be responsible for such discrepancy. A possible explanation is that the strong tidal forces operating in the convective envelope of \hr\ are responsible for its very weak \drot, though observations of other close binaries are obviously needed to confirm this result and disentangle the influence of a tidal torque from that of other stellar parameters. 

The lifetime of surface structures, reported to undergo fast changes over periods as short as a couple of weeks, is similar to values suggested by earlier studies of young dwarfs (Barnes et al. 1998), but much shorter than on \hr\ where typical active regions remain stable on timescales as long as 4 to 6 weeks. The smaller size of most magnetic and spotted regions of \hd\ may partly explain this difference in lifespan, as well as its stronger \drot\ producing a more intense shear of active regions.

\section{Conclusions and prospectives} 
\label{conclusion} 

This study reports the detection of a photospheric magnetic field on the FK~Com giant \hd. The spatial distribution of the field is reconstructed by means of ZDI. The magnetic topology of \hd\ presents at all observing epochs large regions in which field lines are mostly azimuthally oriented. This observation, together with similar ones reported for other fast rotators (Donati et al. 2003a), suggests that the dynamo processes generating the magnetic activity of \hd\ may be active very close to the stellar surface.

Local short-term evolution of surface brightness and magnetic structures are observed on timescales as short as a couple of weeks. A polarity reversal is reported during summer 2002 for high-latitude magnetic regions hosting a radial field. A slow increase of the fraction of magnetic energy stored at low-latitude in the large-scale azimuthal component is also observed. Beside these local evolutions, the largest features of the brightness and magnetic topologies remain stable over several years. Although the observed surface changes remain mostly limited to small structures on our 4.5~yr monitoring, future observations may tell us whether the magnetic field of \hd\ undergoes a cyclical evolution similar to that observed on the Sun.

It is finally suggested that the surface of \hd\ is sheared by a solar-like differential rotation. The difference in rotation rate between equatorial and polar regions is reported to be about 1.5 times that of the Sun.

The observations reported here are at the limit of the capacity of the MuSiCoS spectropolarimeter, in terms of \sn\ and spectral resolution. Much more accurate observations are therefore expected for the same star when the new generation of spectropolarimeters becomes available, with ESPaDOnS at the Canada-France-Hawaii Telescope (Donati 2003) and NARVAL at TBL (Auri\`ere 2003). Better data may in particular allow us to probe more discreet photospheric effects, like possible secular fluctuations in the amount of differential rotation, as already reported for other active stars (Donati et al. 2003b). 

\section*{ACKNOWLEDGMENTS}

PP acknowledges the Portuguese Funda\c c\~ao para a Ci\^encia e a Tecnologia for grant support \# SFRH/BPD/11139/2002. GAW, and JDL acknowledge grant support from the Natural Sciences and Engineering Research Council of Canada (NSERC). JMO acknowledges support of the UK Particle Physics and Astronomy Research Council (PPARC). We thank an anonymous referee whose comments helped to improve the manuscript.

\end{document}